\documentclass[aps,prd,preprint,showpacs,groupedaddress]{revtex4}
\usepackage{graphicx}% Include figure files

\newcommand{\sla}[1]{{#1}\!\!\!/}
\newcommand{\gammu}{\gamma^{\mu}}
\newcommand{\spac}{\ensuremath{\ }}

\newcommand{\hlo}{\ensuremath{h_{\rm L}^0}}
\newcommand{\bhlo}{\ensuremath{{\bar{h}}_{\rm L}^0}}

\newcommand{\Dro}{\ensuremath{D_{\rm R}^0}}
\newcommand{\bDro}{\ensuremath{{\bar{D}}_{\rm R}^0}}
\newcommand{\Dlo}{\ensuremath{D_{\rm L}^0}}
\newcommand{\bDlo}{\ensuremath{{\bar{D}}_{\rm L}^0}}
\newcommand{\lu}{\ensuremath{L^u}}
\newcommand{\ld}{\ensuremath{L^d}}

\newcommand{\ru}{\ensuremath{R^u}}
\newcommand{\rd}{\ensuremath{R^d}}

\newcommand{\lud}{\ensuremath{(L^u)^{\dagger}}}

\newcommand{\Ulo}{\ensuremath{U_{\rm L}^0}}
\newcommand{\Uro}{\ensuremath{U_{\rm R}^0}}
\newcommand{\bUlo}{\ensuremath{\bar{U}_{\rm L}^0}}
\newcommand{\bUro}{\ensuremath{\bar{U}_{\rm R}^0}}

\newcommand{\es}{E$_6$~}

\begin{document}

\preprint{EFI-03-37}
\preprint{hep-ph/0309254}

\title{Exotic $Q = -1/3$ Quark Signatures \\ at High-Energy Hadron Colliders}


\author{Troy C. Andre}
\email[]{troy@hep.uchicago.edu}
%\thanks{}

\author{Jonathan L. Rosner}
\email[]{rosner@hep.uchicago.edu}
%\thanks{}
\affiliation{Enrico Fermi Institute and Department of Physics \\ University of Chicago, Chicago, IL 60637}


\date{\today}

\begin{abstract}
Isosinglet vector-like quarks are predicted by some unified theories
of electroweak and strong interactions.  We study hadron collider
signatures for the production and decay of isosinglet vector-like
quarks with charge $-1/3$.  Previous analyses of
Run I data from the Fermilab Tevatron are used to set lower limits of
100--200 GeV$/c^2$ on the mass of such quarks, depending on
assumptions about mixing with Standard Model quarks and the mass of
the Higgs boson.  For future Tevatron data (${\rm E}_{\rm c.m.} =
1.96$ TeV) the corresponding mass range is ($100$--$270$,
$100$--$320$) GeV/$c^2$ for ($1$, $10$) ${\rm fb}^{-1}$.  At the CERN
Large Hadron Collider (LHC) (${\rm E}_{\rm c.m.} = 14$ TeV, $100$
${\rm fb}^{-1}$), an analysis of flavor-changing neutral-current decay
modes should probe an $h$ quark mass range of $100$--$1100$ GeV/$c^2$.
\end{abstract}

%\keywords{}

\maketitle

\section{\label{sec:intro}Introduction}
The currently known fermions consist of quarks $(u,c,t)$ with charge
$2/3$, quarks $(d,s,b)$ with charge $-1/3$, leptons $(e,\mu,\tau)$
with charge $-1$, and neutrinos $(\nu_e,\nu_\mu,\nu_\tau)$ with charge
$0$.  In the Standard Model (SM), these fermions are arranged into
structures that transform under the gauge group SU(3)$_{\rm
c}\times$SU(2)$_{\rm L}\times$U(1)$_{\rm Y}$.  However, a deeper
understanding of the particle spectrum and its pattern of
charge-changing weak transitions is still unknown.

One may try to understand the fermion spectrum and couplings by
embedding the Standard Model in a larger gauge group.  For example,
unified theories of the electroweak and strong interactions based on
the group SO(10)~\cite{SOten} can accommodate precisely this set of
fermions within three 16-dimensional spinor representations.  Larger
unified groups, like \es \cite{SOten,E6,Rosner:1985hx}, not only
contain the Standard Model fermions but also predict the existence of
new particles.  The discovery of new particles predicted by higher
dimensional gauge theories would provide insight on the organization
of matter into a fundamental theory.

A unified theory based on the gauge group \es is phenomenologically
interesting because it includes an enlarged lepton sector containing
both massive and sterile neutrinos, an enlarged quark sector
containing charge $-1/3$ isosinglet vector-like quarks (ISVLQ), and
additional gauge bosons (e.g. Z$^{\prime}$) \cite{Rosner:1985hx,
  Rosner:rd}.  For additional information on building low-energy
models from the gauge group ${\rm E}_6$, consult
Refs.~\cite{Rosner:1985hx, Rosner:rd, Chang:1986ru, Hewett:1988xc,
Nardi:1992am, Rizzo:1998ut}.

In this paper, we study weak isosinglet vector-like quarks with charge
$-1/3$.  Pair production of ISVLQs at high-energy hadron colliders is
expected to be dominated by quantum chromodynamics (QCD) and thus to
be precisely calculable.  The decays of these particles depend on
their mixing with SM down-type quarks~\cite{Branco:1986my,
Robinett:dz, Campbell:1986xd, Barger:1985nq, delAguila:se,
Langacker:1988ur, Fritzsch:1995nx, Hou:papers}.  We analyze the
prospects for producing and detecting these exotic quarks at the
Fermilab Tevatron and the CERN Large Hadron Collider (LHC).  In
particular, we consider both charged-current (mediated by ${\rm
  W}^{\pm}$ bosons) and flavor-changing neutral-current (mediated by
${\rm Z}^{0}$ and Higgs bosons) decays of the isosinglet vector-like
quark.  The sensitivity of these estimates to assumptions about mixing
between exotic quarks and those of the Standard Model is explored.
Related earlier studies have appeared in Refs.~\cite{Rosner:1985hx,
Branco:1986my, Barger:1985nq}.

The paper is organized as follows.  In Section~\ref{sec:model}, we
review relevant properties of the exotic quarks, and we introduce a
model for their mixing with ordinary quarks.  In
Section~\ref{sec:constraints}, we briefly review constraints on the
new mixing parameters.  Based on these constraints, we propose a
phenomenological parametrization of the Cabibbo-Kobayashi-Maskawa
(CKM) matrix.  Signals at the Fermilab Tevatron and the CERN LHC are
treated in Sections~\ref{sec:pandd} and~\ref{sec:panddLHC},
respectively, while Section~\ref{sec:conclusion} concludes.

% Start ``The Model'' Section
\section{The Model}
\label{sec:model}
\subsection{Matter States Expected in \es}
\label{subsec:matter}
In a unified electroweak theory based on the \es gauge group, the
fundamental ($27$-dimensional) representation contains additional
quarks with charge $-1/3$ and additional charged and neutral leptons.  The
exotic matter content of a single \es family is summarized in
Table~\ref{tab:27plet} \cite{Rosner:rd}.  We assume that there are
three $27$-plets, corresponding to the three lepton-quark families.

\begin{table}
\caption{\label{tab:27plet}Exotic fermions in a 27-plet of \es.}
\begin{ruledtabular}
\begin{tabular}{c c c c c c}
SO(10) & SU(5) &  State  & $Q$   & $I_L$ & $I_{3L}$  \\ \hline
10     & 5     &  $h^c$  & $1/3$ &   0   &   0       \\
       &       &  $E^-$  & $-1$  &  1/2  & $-1/2$    \\
       &       & $\nu_E$ &  0    &  1/2  &   1/2     \\
       & $5^*$ &   $h$   & $-1/3$ &   0   &   0      \\
       &       &  $E^+$  &   1    &  1/2  &   1/2    \\
       &    & $\bar \nu_E$ &   0    &  1/2  & $-1/2$ \\ \hline
1      &   1   &  $n_e$  &   0    &   0   &   0      \\
\end{tabular}
\end{ruledtabular}
\end{table}

We adopt a ``bottom-up'' approach to a three-generation
\es gauge field theory.  Let $M_1,\spac M_2, \spac M_3$ denote the
masses of the three exotic charge $-1/3$ quarks.  For simplicity we
assume that there is a mass hierarchy between the exotic quarks such
that $M_1 \ll M_2,\spac M_3$.  Hence, one of the exotic quarks will
lie closer in the mass spectrum to the SM quark masses.  In accordance
with the literature~\cite{Rosner:1985hx, Robinett:dz, Barger:1985nq},
we denote this exotic quark as $h$; one
should not confuse the exotic $h$ quark with the SM Higgs boson,
denoted ${\rm H}^0$.  We assume $h$ is the dominant exotic quark
(relative to the two other exotic quarks) which mixes with the
down-type SM quarks.  Moreover, we assume that the exotic leptons
(charged and neutral) do not significantly influence SM interactions
or exotic-quark signatures at the center of mass (CM) energy of the
Tevatron or the LHC.  Production of exotic leptons, if kinematically
allowed, should only proceed via the electroweak sector of the theory,
and thus should be suppressed with respect to exotic quark production.
Under these assumptions, we have effectively reduced the \es model at
the CM energies of the Tevatron and the LHC to a model which contains
the SM along with a single down-type exotic quark.

The exotic quark, $h$, is a down-type quark, but unlike the SM quarks,
the ISVLQ is a singlet under the SU$(2)_{\rm L}$ factor of the SM
gauge group.  The singlet nature of the down-type ISVLQ introduces new
mixings and interactions between the quarks.  In the remainder of this
section, we construct these interactions and explore their influence
on the CKM matrix.

In this paper, the SM fermions are labeled by a generation index $i$
($i = 1,2,3$) and we label the $h$ quark by the index value $4$.  The
indices $(i,j,k)$ run from one to three and the indices $(l,m,n)$ run
from one to four.  The ``L'' and ``R'' subscripts are employed to
denote the left- and right-handed components of fermion fields
(i.e., $u_{\rm L,R} = {\rm P}_{\rm L,R} u = [\frac{1}{2}(1\mp
\gamma^5)] u$ in our notation).  To facilitate our discussion it is
useful to define the left-handed quark doublet, and the left- and
right- handed quark field vectors:
\begin{eqnarray}
\begin{tabular}{ccc}
${\mathcal Q}_i = {\left( u_{{\rm L}i},\spac d_{{\rm L}i} \right)}^{\rm T}$, &
$U_A = {\left( u_A,\spac c_A,\spac t_A \right)}^{\rm T}$, & 
$D_A = {\left( d_A,\spac s_A,\spac b_A,\spac h_A \right)}^{\rm T}$,
\end{tabular} 
\end{eqnarray}
where $A = L,R$.  We denote the flavor eigenstates via the ``0'' superscript.

In the flavor basis, the kinetic piece of the quark Lagrangian is
obtained via minimal coupling
\begin{eqnarray}
\label{eqn:lagweak}
{\mathcal L}_K & = & \bar{{\mathcal Q}}_{i}^{0} (i\sla{\partial}) 
{\mathcal Q}_{i}^0 + {\bhlo (i\sla{\partial}) \hlo} + {\bUro (i\sla{\partial}) 
\Uro} + {\bDro (i\sla{\partial}) \Dro} \nonumber \\
 & & + g{W_{\mu}^-}J_{W^+}^{\mu} + g{W_{\mu}^+}J_{W^-}^{\mu} + 
g{Z_{\mu}^0}J_{Z}^{\mu} + e{A_{\mu}}J_{EM}^{\mu}, 
\end{eqnarray}
where the $J_{EM}$, $J_{W^{\pm}}$, and $J_Z$ are the electromagnetic,
charged-weak, and neutral-weak current operators.  In the weak
eigenbasis these currents take the form
\begin{eqnarray}
\label{eqn:intweak}
J_{W^+}^{\mu} & = & \frac{1}{\sqrt{2}}\sum_{i=1}^3 \left[ {\bar{U}^0_{{\rm L}i}}
\gammu {D^0_{{\rm L}i}} \right] \spac , \spac
J_{W^-}^{\mu} = \frac{1}{\sqrt{2}}\sum_{i=1}^3 \left[ {\bar{D}^0_{{\rm L}i}}
\gammu {U^0_{{\rm L}i}} \right] \nonumber \\
J_{Z}^{\mu} & = &
\frac{1}{c_W}\left\{ C_{\rm L}^u \left[ \bar{U}^0_{\rm L} \gammu U^0_{\rm L} \right] + 
C_{\rm R}^u \left[ \bar{U}^0_{\rm R} \gammu U^0_{\rm R} \right] 
+ \sum_{i=1}^3 C_{\rm L}^d \left[ \bar{D}^0_{{\rm L}i} \gammu D^0_{{\rm L}i} \right] + 
C_{\rm R}^d \left[ \bar{D}^0_{\rm R} \gammu D^0_{\rm R} \right] 
+ \, \frac{1}{3} s^2_W \left[ \bhlo \gammu\hlo \right] \right\} \nonumber \\
J_{EM}^{\mu}  & = & \frac{2}{3} \left[ \bUlo \gammu \Ulo + 
\bUro \gammu \Uro \right] - \frac{1}{3} \left[ 
\bDlo \gammu \Dlo + \bDro \gammu \Dro \right],
\end{eqnarray} 
where $s_W$ and $c_W$ are the sine and cosine of the weak mixing
angle, $C_{\rm L}^a = I_{\rm L}^a - Q_a s^2_W$ and $C_{\rm R}^a = -Q_a
s^2_W$ ($a = u,d$).

In the Standard Model, quark masses are generated when the Higgs
doublet acquires a non-zero vacuum expectation value (VEV); the
resulting mass terms in the Lagrangian have $\Delta I_{\rm L} = 1/2$.
Unlike the SM, theories with exotic ${\rm E}_6$ quarks have $\Delta I_{\rm
L} = 0$ mass terms.  In the ISVLQ model, these Dirac mass terms result from
the left-handed $h$ field (recall $h$ is an SU$(2)_{\rm L}$ singlet)
pairing with right-handed quark fields.  We write the quark mass terms
of the Lagrangian in the compact form
\begin{eqnarray}
\label{eqn:lagmass}
{\mathcal L}_{mass} & = & -\bUlo {\mathcal M}^u \Uro - \bDlo {\mathcal M}^d 
\Dro + h.c.,
\end{eqnarray} 
where ${\mathcal M}^u$ and ${\mathcal M}^d$ are the up-type quark and
down-type quark mass matrices, respectively.  The up-type quark mass
matrices mimic those of the SM; however, the down-type quark mass
matrix has been enlarged.  The down-type quark mass matrix contains
both $\Delta I_{\rm L} = 1/2$ and $\Delta I_{\rm L} = 0$ entries.

It is important to distinguish the $\Delta I_{\rm L} = 1/2$ and
$\Delta I_{\rm L} = 0$ mass terms since they may arise from
fundamentally different scales.  The $\Delta I_{\rm L} = 1/2$ mass
terms are derived from spontaneous symmetry breaking (SSB) of the SM
Higgs at the electroweak scale; hence, these mass terms should be
on the order of the electroweak scale or smaller.  On the other hand, the
$\Delta I_{\rm L} = 0$ mass terms do not result from electroweak
symmetry breaking (EWSB); therefore, they may be of any scale,
possibly the unification scale.  We distinguish $\Delta I_{\rm L} =
1/2$ and $\Delta I_{\rm L} = 0$ mass terms in the down-type quark
matrix by lower- and upper-case letters, respectively,
\begin{eqnarray}
\label{eqn:mmatrix}
{\mathcal M}^d & = & \left(
\begin{array}{cccc}
m_{dd} & m_{ds} & m_{db} & m_{dh} \\
m_{sd} & m_{ss} & m_{sb} & m_{sh} \\
m_{bd} & m_{bs} & m_{bb} & m_{bh} \\
M_{hd} & M_{hs} & M_{hb} & M_{hh} 
\end{array}
\right).
\end{eqnarray}

We investigate the production and decay of an exotic \es
quark at the Fermilab Tevatron and the CERN Large Hadron Collider.  To
understand exotic quark decay we must determine the branching
ratios of these exotic fermions to SM particles.  Hence we must
determine the analogue of the CKM matrix
and the no-longer trivial flavor-changing neutral current (FCNC)
matrix for this theory, and we must constrain elements of these
matrices by experimental data.  To determine the CKM and FCNC matrices
we must recast our theory in the {\it mass} eigenbasis.

Each quark mass matrix in this theory may be diagonalized by two
unitary transformations, denoted $L$ and $R$,
\begin{eqnarray}
{\left( {\mathcal M}^d \right)}_{\rm diag} = {\left(\ld\right)}^{\dagger}
{\mathcal M}^d \rd & , & 
{\left( {\mathcal M}^u \right)}_{\rm diag} = {\left(\lu\right)}^{\dagger} 
{\mathcal M}^u\ru 
\end{eqnarray}
where the {\it u} and {\it d} unitary matrices are $3\times 3$ and $4\times 4$, respectively.  The fields in the mass basis are related to the fields in the flavor basis by the $L$ and $R$ transformations:
\begin{eqnarray}
D_L = (\ld)^{\dagger}\Dlo & , & D_R = (\rd)^{\dagger}\Dro \\
U_L = (\lu)^{\dagger}\Ulo & , & U_R = (\ru)^{\dagger}\Uro. \nonumber 
\end{eqnarray}
Using the relations between the mass and flavor eigenstates, we recast
the current operators of Eq. (\ref{eqn:intweak}) in the mass
eigenbasis:
\begin{eqnarray}
J_{W^+}^{\mu} & = & \frac{1}{\sqrt{2}} V_{in} \left[ \bar{U}_i \gammu {\rm P}_{\rm L} D_n \right] \spac , \spac
J_{W^-}^{\mu} = \frac{1}{\sqrt{2}} V_{in}^{*} \left[ \bar{D}_{n} \gammu {\rm P}_{\rm L} U_{i} \right] \\ 
J_Z^{\mu} & = &
\frac{1}{c_W} \left\{ C_{\rm L}^u \left[\bar{U}_i \gammu {\rm P}_{\rm L} U_i\right] + C_{\rm R}^u \left[ \bar{U}_{i} \gammu {\rm P}_{\rm L} U_{i} \right] + X_{nm} \left[ \bar{D}_n \gammu {\rm P}_{\rm L} D_{m} \right] + C_R^d \left[ \bar{D}_{n} \gammu D_{n} \right] \right\} \nonumber \\
J_{EM}^{\mu}  & = & \frac{2}{3} \left[ \bar{U}_i \gammu U_i \right] -\frac{1}{3} \left[ \bar{D}_n \gammu D_n \right], \nonumber
\end{eqnarray} 
where 
\begin{eqnarray}
\label{eqn:ckmfc}
V_{in} = \sum_{k=1}^3 \lud_{ik} \ld_{kn} & , & X_{nm} = C_L^d \delta_{nm} + \frac{1}{2} {\ld}^{*}_{4n}\ld_{4m} ,
\end{eqnarray}
denote the CKM and FCNC matrices ($i=1,2,3$, and $m,n = 1,2,3,4$),
respectively.  One obtains the expression for the FCNC matrix by
transforming the coupling of the left-handed down-type quarks to the
${\rm Z}^0$ from the weak basis to the mass basis,
\begin{eqnarray}
{\mathcal L} & \supset & \frac{g}{c_W} Z^0_{\mu} \left[\bar{D}^0_{{\rm L},m} \gamma^{\mu} \left( C^d_{\rm L}\delta_{mn} + \frac{1}{2}\delta_{4m}\delta_{4n} \right) D^0_{{\rm L},n}\right] \nonumber \\
             & \supset & \frac{g}{c_W} Z^0_{\mu} \left[\bar{D}_{{\rm L},m^{\prime}} \gamma^{\mu} L^{d*}_{m^{\prime}m} \left( C^d_{\rm L}\delta_{mn} + \frac{1}{2}\delta_{4n}\delta_{4m} \right) L^{d}_{nn^{\prime}} D_{{\rm L},n^{\prime}}\right] \nonumber \\
             & \supset & \frac{g}{c_W} Z^0_{\mu} \left[ \bar{D}_{m^{\prime}} \left( C^d_{\rm L}\delta_{m^{\prime}n^{\prime}} + \frac{1}{2}L^{d*}_{4n^{\prime}}L^{d}_{4m^{\prime}} \right) \gamma^{\mu} {\rm P}_{\rm L} D_{n^{\prime}} \right].
\end{eqnarray}
Note: If the ISVLQ were an up-type quark, then the ${\rm Z}^0$ boson would
have a corresponding FCNC coupling to the up-type quarks of the form,
$C^u_{\rm L}\delta_{mn} - \frac{1}{2}L^{u*}_{4n}L^{u}_{4m}$.

This theory differs from the SM in the structures of its CKM and FCNC
matrices.  In the SM, the CKM matrix is a $3\times 3$ {\it unitary}
matrix, while in the ISVLQ model it is a $3\times 4$ matrix
(non-unitary).  Hence the ``unitarity triangle'' approach to determining
CKM parameters must be abandoned in favor of a ``unitarity
quadrangle'' \cite{unitquad}.  In the SM, the FCNC matrix
is diagonal, hence, there are no tree-level flavor-changing neutral
couplings.  In the ISVLQ model this is no longer true; in essence, one
abandons the Glashow-Iliopoulos-Maiani (GIM)
mechanism~\cite{Glashow:gm} which suppresses flavor-changing neutral
currents.  In Section 3, we investigate the structure of the CKM and
FCNC matrices [both matrices are related to the \ld\spac matrix].

\subsection{Feynman Rules}
\label{subsec:feynman}
\begin{figure}
\includegraphics[scale=1.0]{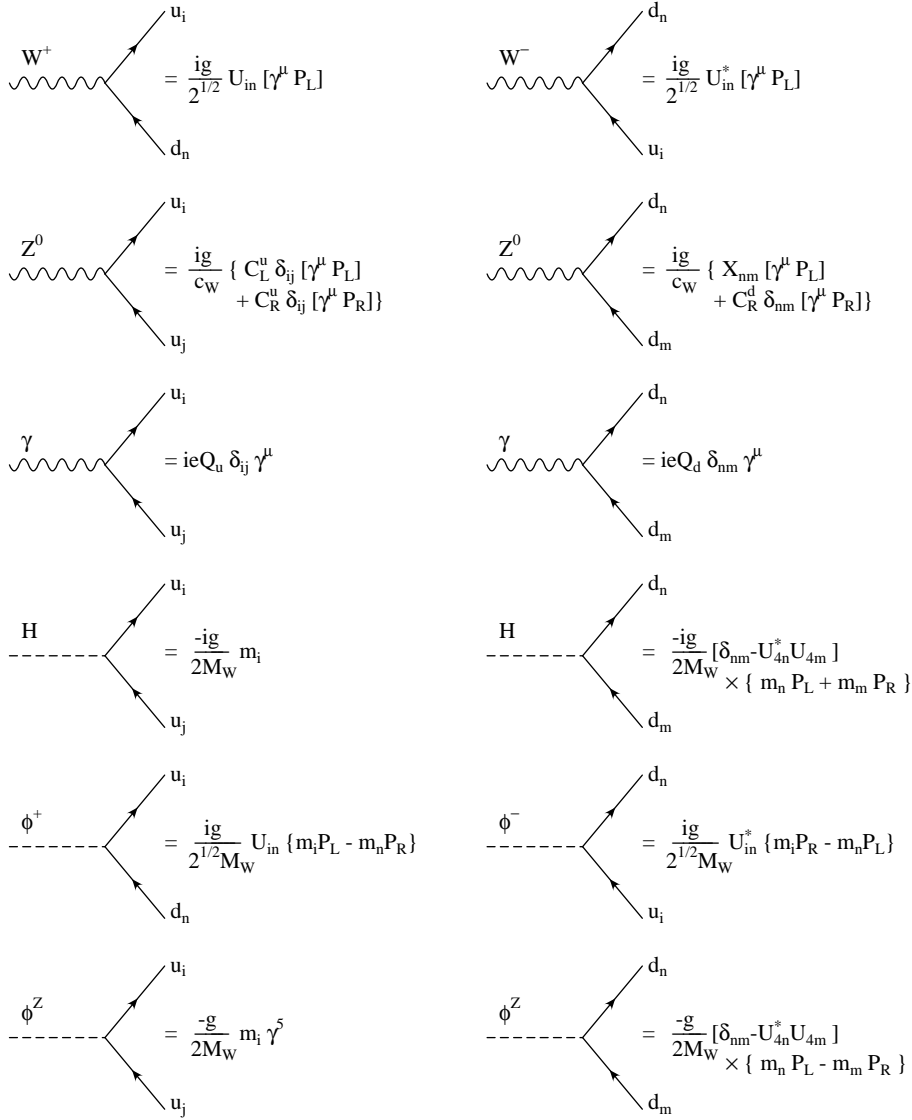}
\caption{\label{fig:feynman}Feynman rules for a model with one ISVLQ
down quark. The indices $i,j = 1,2,3$ and the indices $n,m = 1,2,3,4$.}
\end{figure}

As explained in Section~\ref{subsec:matter}, we consider a simple \es inspired extension
of the SM, in which one down-type ISVLQ interacts with the SM
particles.  In Fig.~\ref{fig:feynman}, we summarize the effects of
this ISVLQ on the quark interactions with SM gauge and scalar fields.
In this paper, all calculations are performed in the Feynman-'t Hooft
gauge.

The addition of an ISVLQ not only induces Z$^0$-mediated tree-level
FCNC interactions, but also induces tree-level FCNCs
mediated by the Higgs, H$^0$, and Goldstone, $\phi^{Z}$, bosons.
Therefore, in certain instances, the branching ratio of the $h$ quark
to a Higgs boson is large.

\section{Constraints and the Parametrization of the CKM Matrix}
\label{sec:constraints}
\subsection{Physical Parameters}
\label{subsec:physparams}
It is useful to parametrize the complex non-unitary $3 \times 4$ CKM
matrix in terms of physical parameters.  The number of physical
parameters of a theory may be determined by analyzing its symmetries.
Consider the up- and down-type quark mass matrices in
Eq. (\ref{eqn:lagmass}); each mass matrix is complex, hence there are
$(9+16)\times 2 = 50$ real parameters.  To determine the number of
physical parameters let us {\it turn off} the mass matrices.  Setting
the quark mass matrices to zero, the theory admits a global quark symmetry,
\begin{eqnarray}
{\mathcal G}_{Global}({\mathcal M}^u = {\mathcal M}^d = 0) & = & {\rm U}(3)_Q \times {\rm U}(4)_{\bar{d}} \times {\rm U}(3)_{\bar{u}} \times {\rm U}(1)_{h},
\end{eqnarray}     
where the subscript $Q$, $\bar{d}$, $\bar{u}$, and $h$ correspond to
the left-handed SM quarks (transforming as $2$ under SU$(2)_{\rm L}$),
right-handed down-type quarks, right-handed up-type quarks, and
left-handed ISVLQ, respectively.  This global symmetry contains
$9+16+9+1=35$ real parameters.  If we turn the quark mass matrices
back on, then the global symmetry is spoiled and we are left with a
remnant global $U(1)$ quark symmetry.  Consequently, there are
$50-35+1 = 16$ {\it physically} significant real parameters: $7$ quark
masses, and $9$ mixing parameters ($6$ angles and $3$ phases).  One
uses the $9$ physical mixing parameters to parametrize the unitary
matrices of the theory (\ld, \rd, \lu, and \ru) and hence the CKM and
FCNC matrices.

Assuming that we have solved for the up-type quark mass eigenstates,
then the CKM matrix expression (\ref{eqn:ckmfc}) reduces to
\begin{eqnarray}
U_{in} & = & {\ld}_{in},  
\end{eqnarray}
where $i=1,2,3$ and $n=1,2,3,4$.  If we parametrize the $4\times
4$ {\it unitary} \ld\spac matrix, we recover the CKM matrix by
restricting to the upper $3\times 4$ sub-matrix.  We recover the
FCNC matrix by selecting appropriate combinations of the $L^d_{4m}$
entries [see Eq.~(\ref{eqn:ckmfc})].

\subsection{Constraints on $h$ Quark Mixings}
\label{subsec:CKMconstraints}
The \ld\spac matrix may be parametrized by nine physical parameters;
however, the ISVLQ model does not specify the magnitude and pattern of
the resulting interactions.  Because we are interested in the
production and decay of $h$ quarks, constraints on the magnitude of
the new mixings between the $h$ quark and the SM quarks are important.
To determine the magnitude and pattern of these and other interactions
(charged-current and flavor-changing neutral-current), experimental
data must be examined.  Using experimental constraints on elements of
the CKM and FCNC matrices, one may infer the structure of the \ld\spac
matrix [consult Eq.~(\ref{eqn:ckmfc})].

Data from precision electroweak experiments and low-energy
flavor-changing neutral processes help to constrain CKM and FCNC
matrix elements.  We use recent constraints on CKM and FCNC
matrix elements obtained by
Refs.~\cite{new_constraint1,new_constraint2}.  For additional
information on constraints of the CKM and FCNC matrix elements in an
ISVLQ model, we refer the reader to studies discussed in
Refs.~\cite{Robinett:dz, Barger:1985nq, Langacker:1988ur, unitquad,
Silverman:constraints, Lavoura:constraints}.  In
Ref.~\cite{new_constraint1}, the following constraints on the
magnitude of \ld\spac matrix elements were obtained: $0 \leq
|L^d_{14}| \leq 0.087$, $0 \leq |L^d_{24}| \leq 0.035$, $0 \leq
|L^d_{34}| \leq 0.041$, and $0.998 \leq |L^d_{33}| \leq 1$.  The
constraint on the magnitude of $L^d_{33}$ is required by the agreement
of $R_b$ with experiment.  Constraints on the magnitude of the
$L^d_{14}$, $L^d_{24}$, and $L^d_{34}$ elements are obtained from the
observables $|\delta m_B|$, $|\delta m_{B_s}|$, $\epsilon$,
$\epsilon^{\prime}/\epsilon$, the branching ratios for $b\rightarrow s
e^+e^-$, $b\rightarrow s\mu^+\mu^-$, $K^+ \rightarrow \pi^+
\nu\bar{\nu}$, $K_L \rightarrow \mu^+ \mu^-$, and the CP asymmetry
$a_{\psi K_s}$.  Ref.~\cite{new_constraint1} also finds that, at
present, there are no restrictions on $h$ quark masses below $1$
TeV/c$^2$.  Using the constraints on the \ld\spac matrix in addition
to constraints on the mixing between SM quarks, we can parametrize the
\ld\spac matrix.

\subsection{Wolfenstein Parametrization}
\label{subsec:wolfen}
There are a number of ways to parametrize the \ld\spac matrix, and
hence the CKM matrix.  We adopt a
Wolfenstein-inspired~\cite{Wolfenstein:1983yz} parametrization.  We
require the elements of the principal
diagonal and the sub-diagonal directly above the principal diagonal to
be real.  In analogy with the phase in the SM Wolfenstein
parametrization, the two new phases of the matrix are assigned to the
$L^d_{14}$ and $L^d_{24}$ elements.  The presence of additional phases
in the ISVLQ model may lead to additional sources of CP violation; for
studies of CP violation arising from the new phases of the \ld\spac
matrix the reader should consult Refs.~\cite{Silverman:cp, branco:cp, Barenboim:1997qx, yamamoto:cp}.

We write $L^d_{14} = A\nu e^{i\omega_2} \lambda^{2+n_{14}}$, and
$L^d_{24} = A\kappa e^{i\omega_2} \lambda^{2+n_{24}}$, where $\omega_2$
and $\omega_3$ are the two new phases, $\nu$ and $\kappa$ ($\nu, \kappa
\sim {\mathcal O}(1)$) are two of the three new angles, the
Wolfenstein ``$A$'' parameter is inserted for convenience, and $n_{14}$
and $n_{24}$ are integers greater than or equal to $0$.  If the $h$ quark
were to mix strongly with either the $u$ or $c$ quarks [i.e., if $L^d_{14}$ or
$L^d_{24}$ were of ${\mathcal O}(\lambda)$], then current SM mixing
contained in the first two rows of the \ld\spac and CKM matrix would
be affected~\cite{footnote:disfavored}.  The third new
angle, $\xi$, parametrizes the mixing of the ISVLQ with the third
generation of quarks.  We parametrize the $L^d_{34}$ matrix elements as
$\xi\lambda^2$ where $\xi \sim {\mathcal O}(1)$.  The magnitude of the
$L^d_{14}$, $L^d_{24}$, and $L^d_{34}$ elements are consistent with
the constraints on the \ld\spac matrix outlined in
Section~\ref{subsec:CKMconstraints} ($\xi \leq 1$ from bound on $L^d_{34}$).

Using this framework, we create a preliminary sketch of the
\ld\spac matrix,
%\begin{singlespace}
\begin{eqnarray}
\ld & = & \left(
\begin{array}{cccc}
1-\frac{1}{2}\lambda^2 +l_{ud}  & \lambda                         & A\mu e^{i\omega_1} \lambda^3  & A\nu e^{i\omega_2} \lambda^{2+n_{14}}    \\
-\lambda +l_{cd}                & 1-\frac{1}{2}\lambda^2 +l_{cs}  & A\lambda^2                    & A\kappa e^{i\omega_3} \lambda^{2+n_{24}} \\
c_{td}\lambda^3 +l_{td}         & c_{ts}\lambda^2 +l_{ts}         & 1-c_{tb}\lambda^4 +l_{tb}     & \xi\lambda^2                \\
L^d_{41}                        & L^d_{42}                        & L^d_{43}                      & 1-c_{4h}\lambda^4+l_{4b} 
\end{array}
\right),
\end{eqnarray}
%\end{singlespace}
\noindent
where the $c_{nm}$ are multiplicative factors of ${\mathcal O}(1)$, and
the $l_{nm}$ are {\it higher-order} contributions in $\lambda$ to the
$L^d_{nm}$ elements ({\it e.g.} in the $L^d_{32}$ element,
$l_{ts}\sim{\mathcal O}(\lambda^3)$ since $c_{ts}\lambda^2$ is of
second order in $\lambda$).  Using the unitarity of the \ld\spac matrix and
additional assumptions regarding the size of particular elements,
one may obtain a fully parametrized expression for the \ld\spac matrix.

As previously mentioned, we expect the $h$ quark to mix predominately
with the third generation.  Therefore, to simplify the parameter space
we assume that the mixing of the $h$ quark with the first two
generations is small.  Namely, we assume that $n_{14}$ and $n_{24}$
are greater than the order of the $L^d$ matrix parametrization.
Parametrizing $L^d$ to ${\mathcal O}(\lambda^5)$, one obtains the
following matrix,
%\begin{singlespace}
\begin{eqnarray}
\label{eqn:ldmatrix}
\ld = \left(
\begin{array}{cccc}
1-\frac{1}{2}\lambda^2-\frac{1}{8}\lambda^4     & \lambda 
& A\mu e^{i\omega_1}\lambda^3                   & 0                            \\
-\lambda                                        & 1-\frac{1}{2}\lambda^2-\left(\frac{1}{8} +\frac{1}{2}A^2 \right)\lambda^4
& A\lambda^2                                    & 0                            \\
A\left( 1-\mu e^{-i\omega_1}\right) \lambda^3   & -A\lambda^2 - A\left( -\frac{1}{2} + \mu e^{-i\omega_1}\right)\lambda^4  
& 1-\left(\frac{1}{2}\xi^2+\frac{1}{2}A^2\right)\lambda^4   & \xi\lambda^2        \\
0                                               & A\xi\lambda^4 
& -\xi\lambda^2                                 & 1-\frac{1}{2}\xi^2\lambda^4 
\end{array}
\right). 
\end{eqnarray}
%\end{singlespace}
\noindent
In this parametrization of the \ld\spac matrix, the phase $\omega_1 =
-\gamma$ and the angle $\mu = \sqrt{\rho^2 + \eta^2}$.

Using this form of the \ld\spac matrix and Eq.~(\ref{eqn:ckmfc}), the
dominant {\it new} charged-current coupling is between the top and
$h$ quarks.  The strength of the $t$-$h$ charged-current interaction
is $g (2^{-1/2}) \xi\lambda^2$.  In the ISVLQ model,
tree-level FCNCs are expected between all four down-type quarks; however, we
find that tree-level flavor-changing neutral currents between SM
quarks are disfavored (${\mathcal O}$($\lambda^6$)).  Interactions
between the $h$ quark and SM down-type quarks may be sizable; in fact,
interactions between the $h$ quark and $b$ and $s$ quarks are of
${\mathcal O}$($\lambda^2$) and ${\mathcal O}$($\lambda^4$),
respectively.  In particular, the $h$-$b$ interaction is {\it
left-handed} with strength $g(2c_{\rm W})^{-1}\xi\lambda^2$ [to
  ${\mathcal O}$($\lambda^5$) in the parametrization].

The form of the \ld\spac matrix in Eq.~(\ref{eqn:ldmatrix}) is just one
of many possibilities arising from the constraints in
Section~\ref{subsec:CKMconstraints}.  For example, the strength of the
charged-current interaction between the $h$ quark and the $u$ or $c$
quarks may not be negligible ({\it e.g.} $n_{14}$ or $n_{24}$ may
equal zero).  If these charged-current interactions are
relevant, then they will contribute to the ``$td$'' and ``$ts$'' elements of
the CKM matrix.  These additional charged-current interactions lead to
corrections to the $l_{td}$ and $l_{ts}$ terms of the $L^d_{31}$ and
$L^d_{32}$ elements.  In the remainder of this paper, we use the
\ld\spac matrix in Eq.~(\ref{eqn:ldmatrix}).

%%%%%%%%%%%%%%%%%%%%%%%%%%%%%%%%%%%%%%%% Pair Production  %%%%%%%%%%%%%%%%%%%%%%%%%%%%%%%%%%%%%%%%%%%%
\section{$h\bar{h}$ Production and Decay Signatures at Hadron Colliders}
\label{sec:pandd}
In this section we investigate the prospects for $h$ quark observation
at the Fermilab Tevatron and the CERN Large Hadron Collider.  In
$p\bar{p}$ and $pp$ collisions, $h\bar{h}$ production proceeds
predominantly through QCD interactions; therefore, in the remainder of
our analysis, we suppress contributions to $h\bar{h}$ production from
electroweak processes.

\subsection{$h\bar{h}$ Production}
\label{subsec:production}
\begin{figure}
\begin{center}
\begin{tabular}{c}
\resizebox{100mm}{!}{\includegraphics{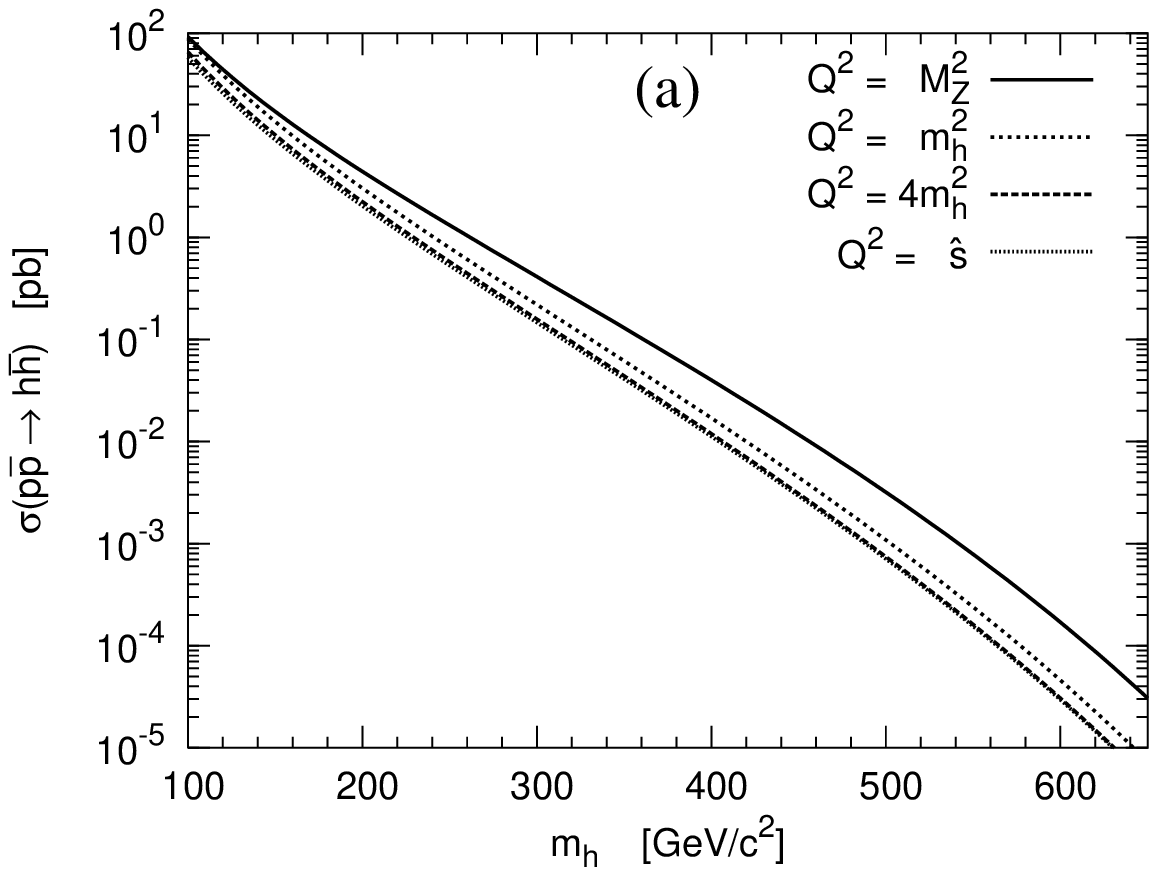}} \\
\resizebox{100mm}{!}{\includegraphics{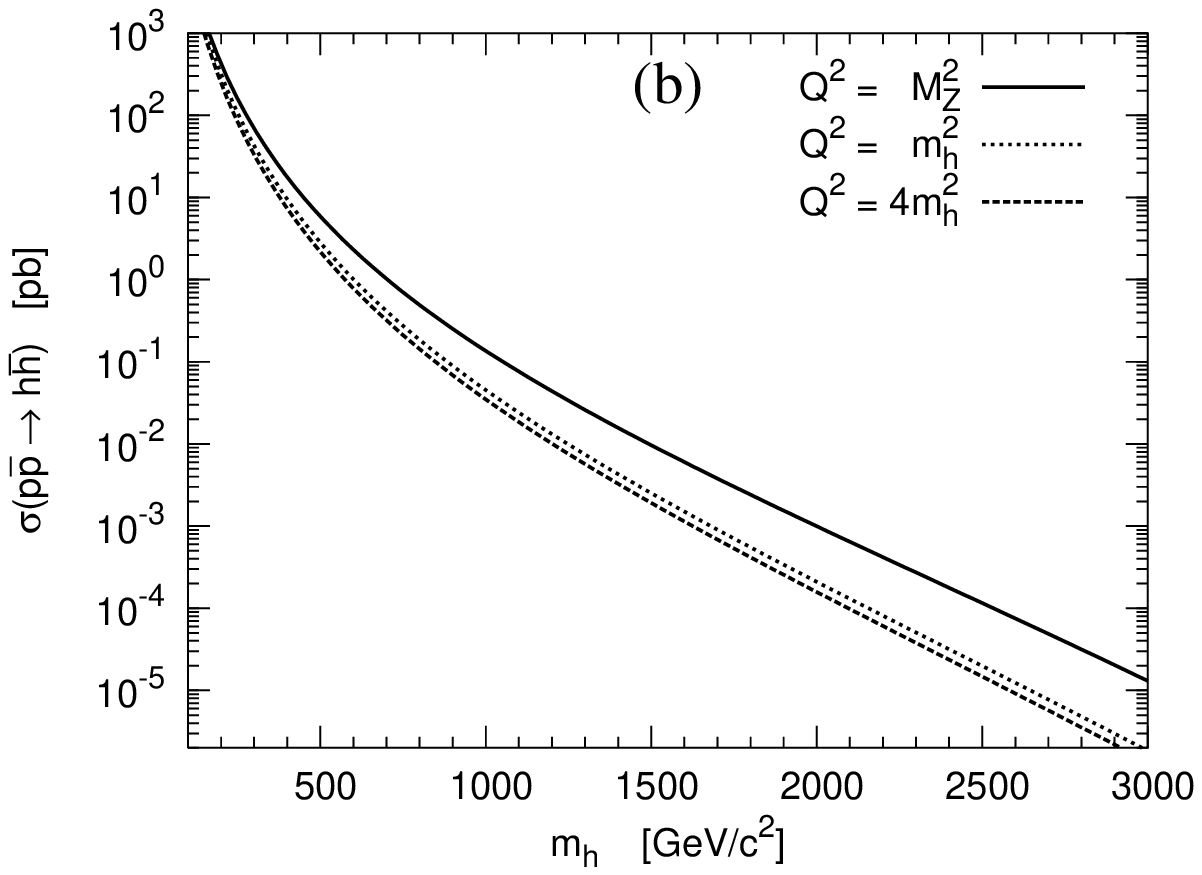}}
\end{tabular}
\caption{\label{fig:xhhb}Plot of the $h\bar{h}$ production cross
section as a function of the $h$ quark mass at {\bf (a)} the Fermilab
Tevatron, $\sqrt{s} = 1.96$ TeV, and at {\bf (b)} the CERN LHC,
$\sqrt{s} = 14$ TeV. The curves in each of these plots correspond to
different choices of the QCD scale ($Q^2$ = $M_Z^2$, $m_h^2$,
$(2m_h)^2$, and $\hat{s}$), where $\hat{s}$ is the square of the
subprocess CM energy.  In the Tevatron plot, the curves for
$Q^2=(2m_h)^2$ and $Q^2=\hat{s}$ overlap, while in the LHC plot we
omit the $Q^2=\hat{s}$ curve since the CTEQ5L structure
functions~\cite{Lai:1999wy} are not defined beyond $10$ TeV.}
\end{center}
\end{figure}

To obtain a basic understanding of the $h\bar{h}$ production rate and
the $h$ quark mass reach at hadron colliders, we plot [see
Fig.~\ref{fig:xhhb}] the tree-level $h\bar{h}$ pair production cross
section at the Fermilab Tevatron ($\sqrt{s} = 1.96$ TeV) and CERN LHC
($\sqrt{s} = 14$ TeV) as a function of the $h$ quark mass, calculated
using the CTEQ5L structure functions~\cite{Lai:1999wy}.  The curves
in each of these plots correspond to different choices of the QCD $Q$
scales ($Q^2$ = $M_Z^2$, $m_h^2$, $(2m_h)^2$, and $\hat{s}$).  For
small $h$ quark masses the cross sections for each
of the $Q$ scales are comparable; however, at $h$ quark masses above
$\sim$ $200$ GeV/c$^2$ there is a large discrepancy between the cross
section with the fixed $Q$ scale ($Q^2$ = $M_Z^2$) and the dynamic $Q$
scale ($Q^2$ = $m_h^2$, $(2m_h)^2$, and $\hat{s}$).  For our analysis
at the Tevatron we use a $Q$ scale which is set by the CM energy of
the subprocess, and at the LHC we use a $Q$ scale equal to twice the
$h$ quark mass.

Using Fig.~\ref{fig:xhhb} and projected integrated luminosities, a
`back of the envelope' upper limit for the $h$ quark mass reach is obtained.
For an integrated luminosity of ($1$, $10$) fb$^{-1}$ at the Tevatron, the
largest $h$ quark mass reachable (producing at least one $h\bar{h}$
event) is $\sim$ ($490$, $550$) GeV/c$^2$.  At a luminosity of $100$
fb$^{-1}$ at the LHC, the largest accessible $h$ quark mass reachable
is $\sim$ 2500 GeV/c$^2$.  These estimates assume perfect event
detection and they assume there are no complications arising from the
decay of the $h$ quarks.

Though Fig.~\ref{fig:xhhb} is instructive, one must take into
consideration final-state signatures and backgrounds to determine a realistic
$h$ quark mass reach.  The $h$
quark is expected to be an unstable particle and hence it will decay
to SM particles.  The three dominant $h$ quark decay processes are $h
\rightarrow t W^-$, $h \rightarrow b Z^0$, and $h \rightarrow b H^0$.
Since each decay mode of the $h$ quark depends on the new mixing
parameter $\xi$, the cross section for $h$ quark production and decay
must also depend on the the mixing parameter.

We calculate the cross sections for $h\bar{h}$ pair
production in conjunction with decay of the $h$ quark into a SM quarks
and gauge bosons.  To facilitate this analysis we use the
CompHEP~\cite{Pukhov:1999gg} software package to calculate the
tree-level cross sections for $h\bar{h}$ production and decay at the
two collider facilities.  For the remainder of this section we shall
restrict our analysis to the Fermilab Tevatron, reserving a brief
discussion of the LHC for Section~\ref{sec:panddLHC}.

\subsection{$h$ Quark Width}
\label{subsec:width}
\begin{figure}
\begin{tabular}{c}
\resizebox{100mm}{!}{\includegraphics{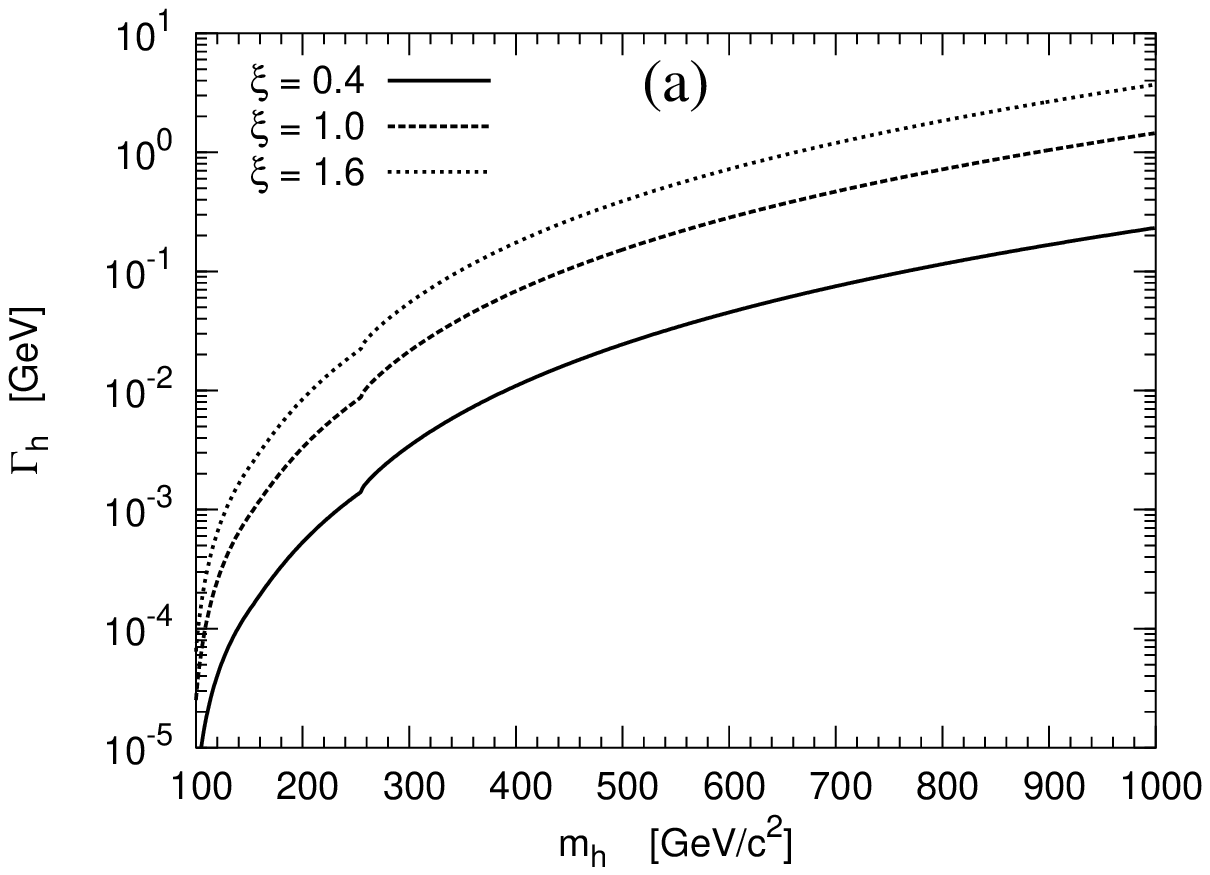}} \\
\resizebox{100mm}{!}{\includegraphics{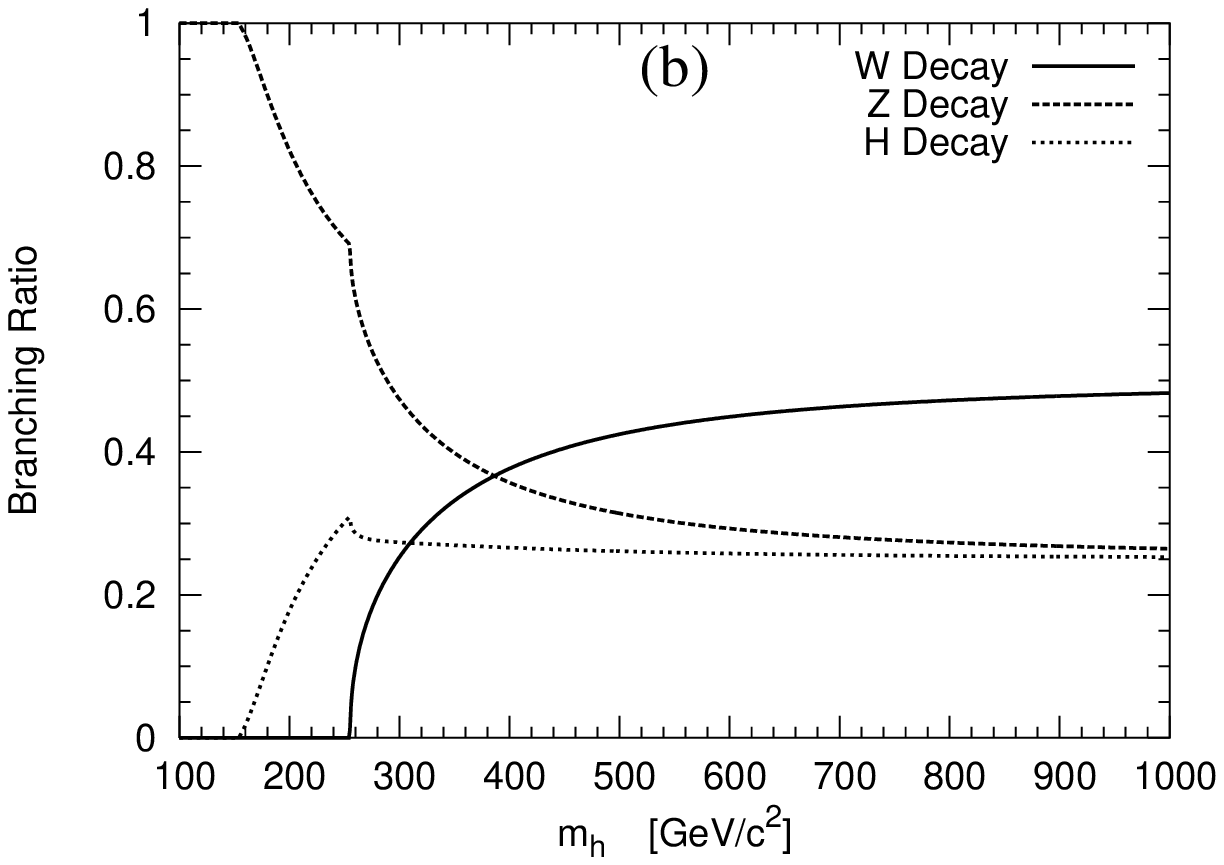}}
\end{tabular}
\caption{\label{fig:hwid}{\bf (a)} Plot of the $h$ quark partial
widths as a function of the mixing parameter $\xi$
($\xi=0.4,1.0,1.6$) and $h$ quark mass, $m_h$. {\bf (b)} Plot of the
$h$ quark branching ratio for a fixed mixing parameter value $\xi=1$
and a variable $h$ quark mass, $m_h$.  In each of these plots, the
Higgs boson mass is taken to be $150$ GeV/c$^2$.}
\end{figure}

To calculate the cross sections for $h\bar{h}$ production and
subsequent decay, we must first calculate the width of the $h$ quark.
The total width of the $h$ quark is the sum of the $h$ quark partial
widths; the decay processes contributing to the partial widths may be
separated into three (decay) channels: $h \rightarrow u_i \; {\rm
W}^-$, $h \rightarrow d_i \; {\rm Z}^0$, and $h \rightarrow d_i \;
{\rm H}^0$.  To leading order the partial widths for the three decay
channels are given by the expressions,
\begin{eqnarray}
\Gamma(h\rightarrow u_i \; {\rm W}^-) & = & \left(\frac{G_F}{4 \sqrt{2}\pi} \right) \left( \frac{c_W^2 M_Z^2}{m_h^2} \right) |L_{i4}|^2 p^*(m_h, m_i, c_W^2 M_Z) F(m_h^2, m_i^2, c_W^2 M_Z^2) \nonumber \\ 
\Gamma(h\rightarrow d_i \; {\rm Z}^0) & = & \left(\frac{G_F}{8 \sqrt{2}\pi} \right) \left( \frac{M_Z^2}{m_h^2} \right) |L^d_{4i}|^2 |L^d_{44}|^2 p^*(m_h, m_i, M_Z) F(m_h^2, m_i^2, M_Z^2) \nonumber \\ 
\Gamma(h\rightarrow d_i \; {\rm H}^0) & = & \left(\frac{G_F}{8 \sqrt{2}\pi} \right) \left( \frac{c_W^2 M_Z^2}{m_h^2} \right) |L^d_{4i}|^2 |L^d_{44}|^2 p^*(m_h, m_i, {\rm M}_{\rm H}) G(m_h^2, m_i^2, {\rm M}_{\rm H}^2), 
\end{eqnarray}
where $p^*$ is the center of mass momentum, and
\begin{eqnarray}
F(m_1^2, m_2^2, M^2) & = & m_1^2 + m_2^2 - M^2 + \frac{1}{M^2}\left(m_1^2 - m_2^2 - M^2\right)\left(m_1^2 - m_2^2 + M^2\right) \nonumber \\
G(m_1^2, m_2^2, M^2) & = & \frac{1}{c_W^2 M_Z^2} \left[ 4 m_1^2 m_2^2 + \left( m^2_1 + m^2_2 \right)\left( m_1^2 + m_2^2 - M^2 \right) \right].
\end{eqnarray}
The center of mass momentum may be expressed as $p^*(m_1, m_2, M) =
\frac{1}{2m_1}{\left(\lambda(m_1^2,m_2^2,M^2)\right)}^{\frac{1}{2}}$,
where $\lambda(a,b,c) = a^2 + b^2 + c^2 -2ab -2ac -2bc$.  The partial
widths for $h$ quark decays (via the W$^-$, Z$^0$ and H$^0$ channels)
are shown in Fig.~\ref{fig:hwid}(a).

In Fig.~\ref{fig:hwid}(b), we plot the branching ratio for the $h$ quark
decaying via Z$^0$, W$^{-}$, and H$^0$ channels.  In this plot the
mixing parameter, $\xi$, is $1$, the mass of the Higgs Boson is
$150$ GeV/c$^2$, and the mass of the $h$ quark is varied from 100 to
1000 GeV/c$^2$.  The ${\rm Z}^{0}$ decay mode is the dominant decay
channel for $h$ quark masses below $200$ GeV/c$^2$; however, the
branching ratio to this mode quickly diminishes as the ${\rm H}^0$ and
${\rm W}^{\pm}$ decay modes become kinematically accessible.  Since the
charged-current $u$ and $c$ couplings to the $h$ quark are small [see
Eq.~(\ref{eqn:ldmatrix})], the branching ratio of the $h$ quark to
W$^-$ is suppressed until $m_h \ge M_W + m_t$.  In the large $h$ quark
mass limit, the ratios of the Z$^0$, H$^0$, and W$^-$ partial widths are
$|L^d_{44}|^2$:$|L^d_{44}|^2$:2.  Using the \ld\spac matrix
paramterization in Eq.~(\ref{eqn:ldmatrix}), these ratios are
$(1-\xi^2\lambda^4):(1-\xi^2\lambda^4):2$.

\subsection{Constraints on ISVLQ model from b$^{\prime}$ searches}
\label{subsec:bpsearches}
Before we discuss specific signatures of $h\bar{h}$ pair production
and their corresponding cross sections, we investigate limits on the
$h$ quark mass from previous experiments.  In particular, we determine
implied limits from the first run of the Fermilab Tevatron.

In Run IB of the Fermilab Tevatron ($\sqrt{s}=1.8$ TeV), the CDF and
D\O\spac collaborations acquired $86.47$ pb$^{-1}$ and $84.5$ pb$^{-1}$ of
data, respectively~\cite{lumin}.  Though a specific analysis of the
ISVLQ model using the Run IB data was not considered, we can infer
limits on the $h$ quark mass from $b^{\prime}$
searches~\cite{Abachi:1996fs, Affolder:1999bs, Abe:1998ee}.  Unlike
the ISVLQ $h$ quark, the $b^{\prime}$ is the charge $-1/3$ member of a fourth quark generation,
$(t^{\prime} \, b^{\prime})^{\rm T}$.  Since the $b^{\prime}$ is a
member of a doublet, the GIM mechanism is preserved and, therefore,
flavor-changing neutral decays of the $b^{\prime}$ are forbidden at
tree-level.  Consequently, flavor-changing $b^{\prime}$ decays occur
via higher-order interactions; they are expected to be suppressed
relative to tree-level interactions.

In these $b^{\prime}$ studies, the CDF and D\O\spac collaborations searched
for a fourth-generation $b^{\prime}$ quark by looking for events where
the $b^{\prime}$ decays via a flavor-changing neutral interaction.  In
particular, both the CDF and the D\O\spac collaborations searched for a
$b^{\prime}$ quark in the following mass regions: (1) $m_{b^{\prime}}
< M_Z$ [see Ref.~\cite{Abachi:1996fs}] and (2) $m_{b^{\prime}} > M_Z + m_b$ but
$m_{b^{\prime}} < m_t$ and $m_{b^{\prime}} < m_{t^{\prime}}$ [see
Ref.~\cite{Affolder:1999bs}].  In the former mass region, the $b^{\prime}$ decays to
a photon and a SM down-type quark, while in the latter mass region the
$b^{\prime}$ decays predominately to a ${\rm Z}^0$ boson and a SM
down-type quark.  Note that in the second mass region,
tree-level charged-current decays to light, up-type quarks are
present; however, they are doubly Cabibbo-suppressed by the small
coupling between the $b^{\prime}$ quark and the light up-type quarks.

\begin{figure}
\begin{tabular}{c}
\resizebox{100mm}{!}{\includegraphics{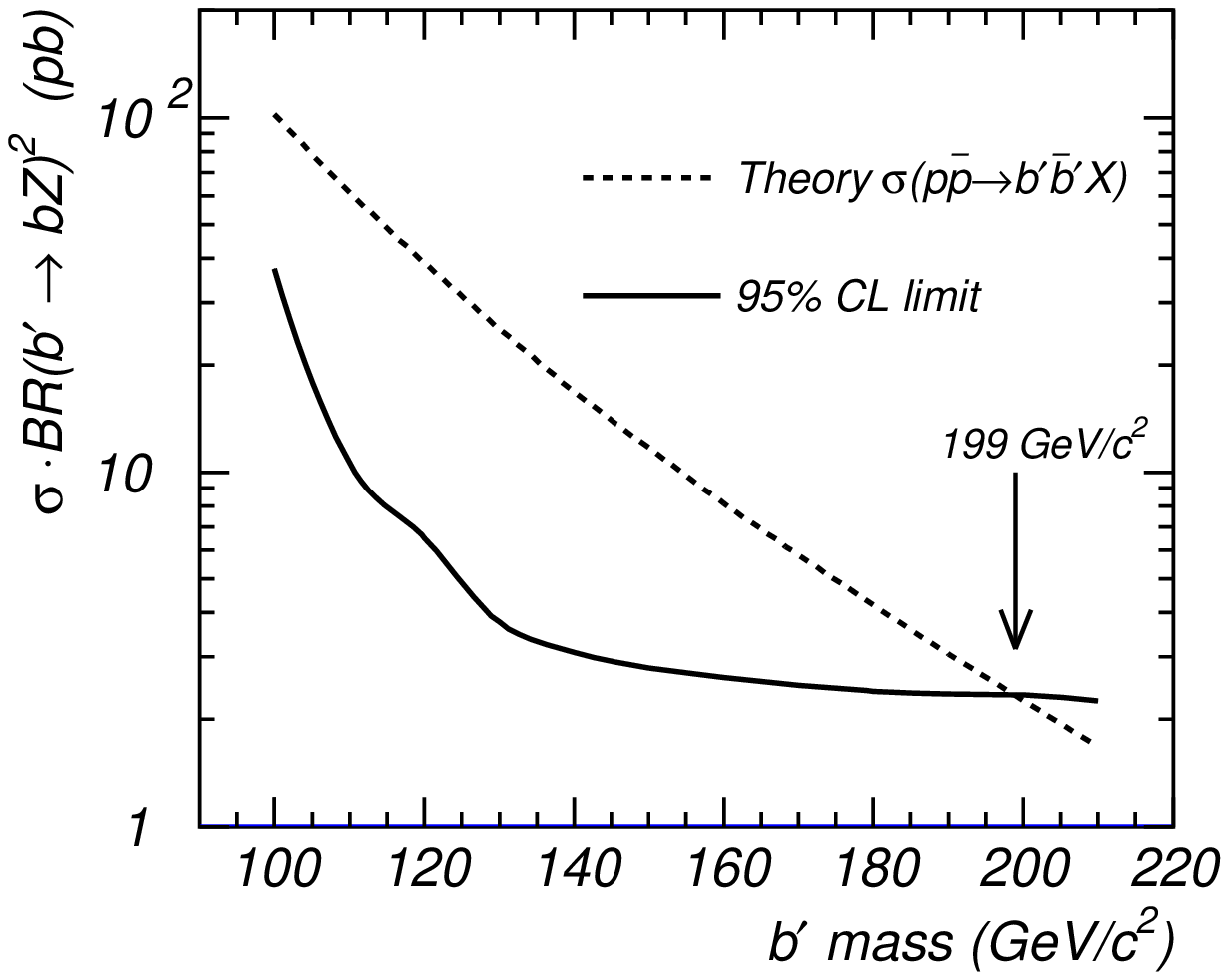}} \\
\resizebox{100mm}{!}{\includegraphics{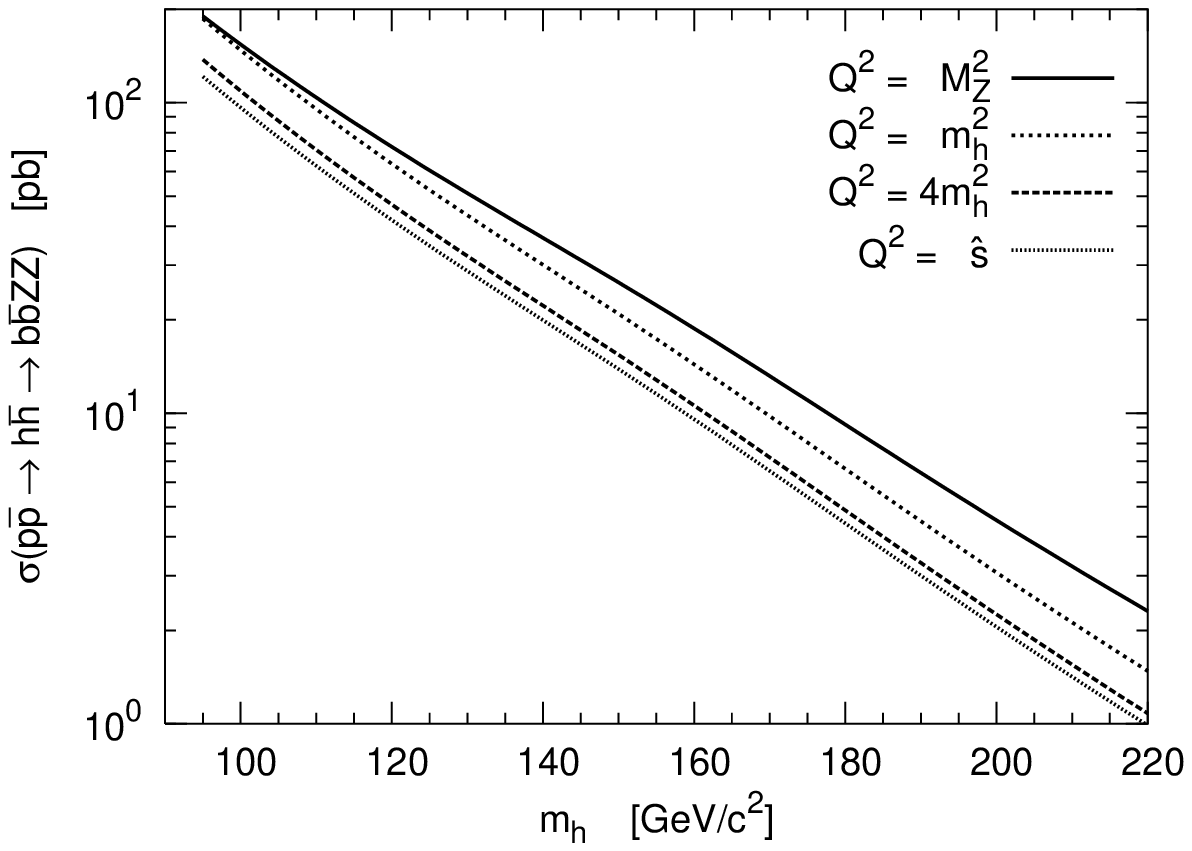}}
\end{tabular}
\caption{\label{fig:run1reach}{\bf (Top)} Exclusion plot for the
$b^{\prime}$ search from run IB of CDF~\cite{Affolder:1999bs}.  {\bf
(Bottom)} Plot of the cross section for $h$ quark pair production and
subsequent decay into two $b$ quarks and two ${\rm Z}^0$ bosons.  In
this plot the mixing parameter $\xi=1$, the Higgs mass is $150$
GeV/c$^2$, and the curves correspond to different values of the $Q^2$ scale.}
\end{figure}

In this paper, we are interested in an ISVLQ $h$ quark with a mass
greater than that of the Z$^0$ boson.  In the top panel of
Fig.~\ref{fig:run1reach}, we reproduce the 95\% confidence limit
(solid line) on the cross section for a $b^{\prime}$ quark of
Ref.~\cite{Affolder:1999bs}.  In the bottom panel of Fig.~\ref{fig:run1reach}, we
plot the cross section for the $h\bar{h}$ production multiplied by the
branching ratio for each $h$ quark to the Z$^0$ mode.  The four curves
in this plot correspond to the aforementioned choices of the QCD $Q$
scale.  Using the $\sqrt{\hat{s}}$ $Q$ scale, one finds that for a
mixing parameter of $\xi=1$ and a Higgs mass of $150$ GeV/c$^2$ the
Run I data exclude a $h$ quark at the 95\% confidence level in the
$100$ -- $200$ GeV/c$^2$ mass range.  For a Higgs mass of $115$
GeV/c$^2$, the Run I analyses imply that an $h$ quark is excluded in
the mass range of $100$ -- $185$ GeV/c$^2$. In the following sections,
we will discuss the effect of the $\xi$ mixing parameter on the $h$
quark mass reach of current and future hadron colliders.

\subsection{$h\bar{h}$ Production and Decay at the Tevatron}
\label{subsec:panddtev}
At the CM energy of the Fermilab Tevatron ($\sqrt{s} = 1.96$ TeV),
$h\bar{h}$ pair production is dominated by subprocesses in which a
quark from a proton and an anti-quark from the anti-proton annihilate
via the strong interaction.  Contributions from subprocesses involved
in gluon-gluon fusion are suppressed by the gluon density of the
proton, and sea quark contributions are suppressed by the sea quark
distributions of the proton.

Once an $h\bar{h}$ pair is produced, each isosinglet quark decays to
a SM quark and an associated gauge or scalar boson.  Though subsequent
decay and hadronization of these particles is expected, we do not
explicitly generate the matrix elements for these individual
processes.  Rather, we generate the matrix elements for
$h\bar{h}$ pair production and the primary decay of the $h$ quarks
into a heavy quark (bottom or top) and an associated gauge or scalar
boson.  We {\it do not} consider processes in which the primary decay
of the $h$ quark results in a light quark.  Processes in which the $h$
quark decays to a W$^-$ and a light quark are disfavored by our choice
of CKM matrix, [see Eq.~(\ref{eqn:ldmatrix})]; moreover, processes in which
the $h$ quark decays to a Z$^0$ or a H$^0$ and a light down-type quark
often lead to complicated, multi-jet event topologies.

For the Tevatron, the $h\bar{h}$ production and decay schemes
considered in this paper are:
%\begin{singlespace}\begin{center}
\begin{center}
\begin{tabular}{lccccl}
1. & $p\bar{p}$ & $\rightarrow$ & $h\bar{h}$ & $\rightarrow$ & $t\bar{t}$ ${\rm W}^+$${\rm W}^-$, \\
2. & $p\bar{p}$ & $\rightarrow$ & $h\bar{h}$ & $\rightarrow$ & $b\bar{b}$ ${\rm Z}^0$${\rm Z}^0$, \\
3. & $p\bar{p}$ & $\rightarrow$ & $h\bar{h}$ & $\rightarrow$ & $b\bar{b}$ ${\rm H}^0$${\rm H}^0$, \\
4. & $p\bar{p}$ & $\rightarrow$ & $h\bar{h}$ & $\rightarrow$ & $b\bar{b}$ ${\rm H}^0$${\rm Z}^0$, \\
5. & $p\bar{p}$ & $\rightarrow$ & $h\bar{h}$ & $\rightarrow$ & $t\bar{b}$ ${\rm W}^-$${\rm Z}^0$ + $\bar{t}b$ ${\rm W}^+$${\rm Z}^0$, \\
6. & $p\bar{p}$ & $\rightarrow$ & $h\bar{h}$ & $\rightarrow$ & $t\bar{b}$ ${\rm W}^-$${\rm H}^0$ + $\bar{t}b$ ${\rm W}^-$${\rm H}^0$.
\end{tabular}
%\end{center}\end{singlespace}
\end{center}
\noindent
Promising final-state signatures and backgrounds for these processes
are discussed in Sections~\ref{subsec:signatures}
and~\ref{subsec:backgrounds}, respectively.  For the time being, we are
interested in the cross sections for these six $h\bar{h}$
production/decay schemes and which, if any, of these schemes provides
the best channel(s) for an $h$ quark search at the Tevatron.
\begin{table}[h]
\begin{ruledtabular}
\caption{\label{tab:tevcuts} We require that the $h$ quark decay
products pass basic pseudorapidity ($\eta$), transverse momentum
($p_T$), and angular separation ($\Delta$R) cuts.  Note that the
angular separation cut is only applied to the bottom quarks and not to
the top quark or the bosons (gauge nor scalar).}
\begin{tabular}{ c c c }
Parameter & Minimum Value & Max Value \\ \hline
$\eta_{\rm b}$            &  $-3.0$  &  $3.0$ \\ 
$\Delta$R($b_1$,\,$b_2$)  &  $ 0.4$  &  --    \\ 
$p^b_T$                   &  $25.0$  &  --    \\ 
\end{tabular}
\end{ruledtabular}
\end{table}

To begin, we impose a loose set of cuts to ensure that the produced
$h\bar{h}$ events conform to {\it basic} geometry and event selection
requirements of the detectors.  These cuts are summarized in
Table~\ref{tab:tevcuts}.  We impose a loose cut on the pseudorapidity
$\eta$ of the bottom quark, $|\eta_b|<3$.  Pseudorapidity is defined
as $\eta = -\log \tan(\theta/2)$, where $\theta$ is the angle between
the particle being considered and the undeflected beam.  In addition to the pseudorapidity cut,
we impose a jet separation cut, $\Delta R = \sqrt{(\Delta \eta)^2 +
  (\Delta \phi)^2}$, to ensure that there is adequate jet separation
for detection.  Finally, we impose a cut on the transverse momentum,
$p_T$, of the bottom quark.  Because the bottom quark is a decay product
of a much heavier $h$ quark, one expects the bottom quarks to be
``hard'' (high momentum) and to have substantial transverse momentum which scales with the mass
of the $h$ quark.  Moreover, because previous data from the Fermilab
Tevatron appear to exclude an $h$ quark up to $\sim 200$ GeV/c$^2$
(when $M_H = 150$ GeV/c$^2$), a ``hard'' cut on the transverse momentum of the $b$ quark will merely
reduce the $h$ quark signal in a previously excluded region ({\it
i.e.} $100$ -- $200$ GeV/c$^2$).  Therefore, we impose a lower bound
of $25$ GeV/c on the transverse momentum of each $b$ quark.  As we shall discuss in Section~\ref{subsec:backgrounds}, the $p_T$ cut and the
$\Delta R$ cut on the $b$ quarks help to reduce backgrounds for these
$h\bar{h}$ pair production processes.  The top quark and the gauge
bosons are unstable particles; therefore, we do not impose any constraints
on the these particles.  When we discuss event signatures in
Section~\ref{subsec:signatures}, we will impose cuts on the decay
products of these particles.

\begin{figure}
\begin{tabular}{c}
\resizebox{100mm}{!}{\includegraphics{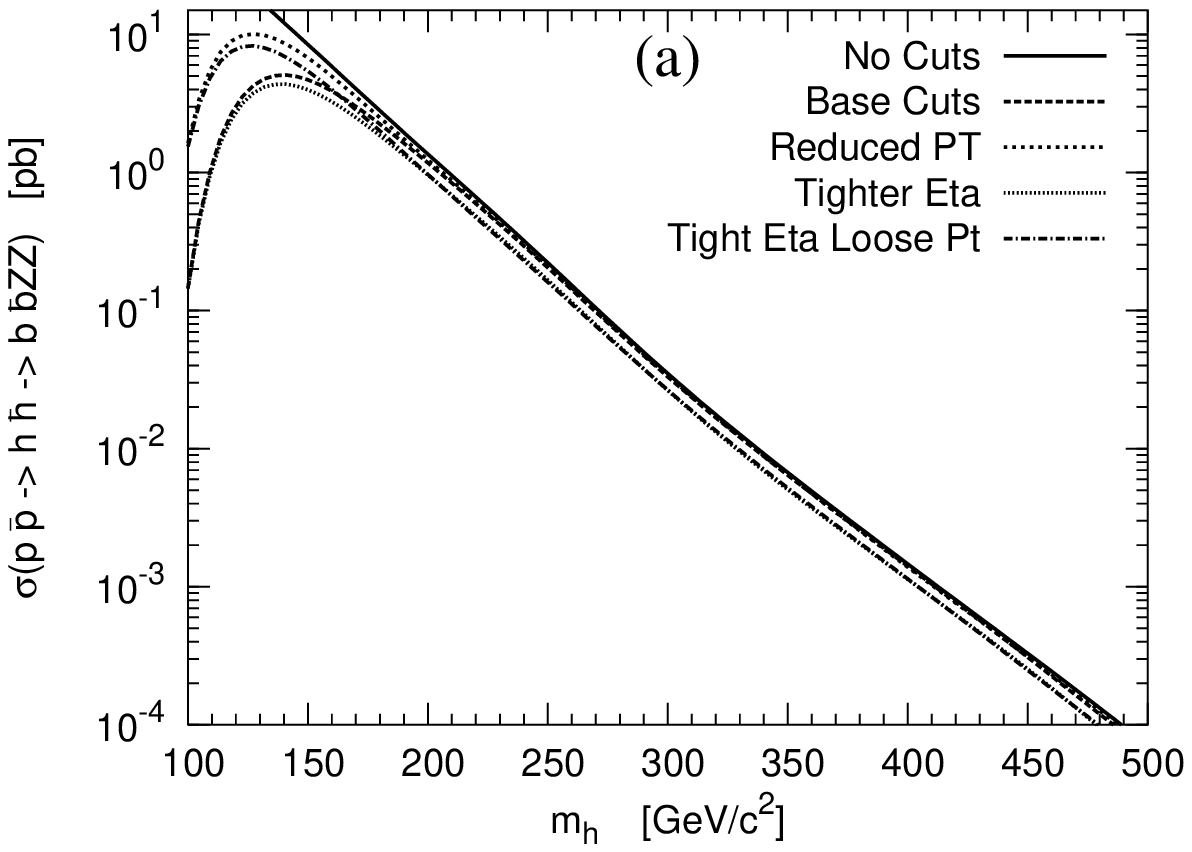}} \\
\resizebox{100mm}{!}{\includegraphics{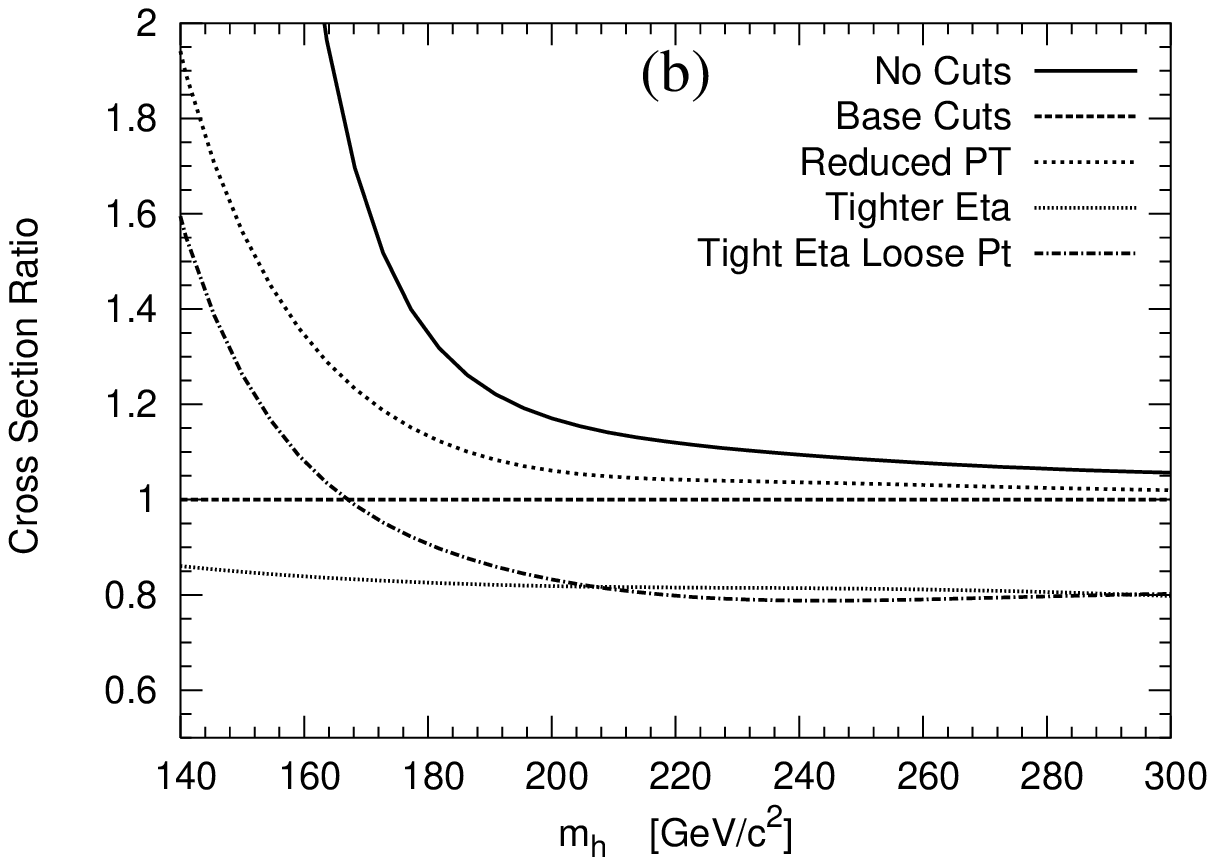}}
\end{tabular}
\caption{\label{fig:cuts}Effect of cuts on the $p\bar{p} \rightarrow
h\bar{h} \rightarrow b\bar{b}{\rm Z}^0{\rm Z}^0$ cross section where
the mixing parameter $\xi = 1$, and the Higgs mass $M_H = 150$
GeV/c$^2$.  {\bf (a)} Plot of $\sigma(p\bar{p} \rightarrow h\bar{h}
\rightarrow b\bar{h}{\rm Z}^0{\rm Z}^0)$ for five different choices of
$b$ quark cuts: (solid) {\it No Cuts}; (long dash) {\it Base Cuts}, see
Table~\ref{tab:tevcuts}; (short dash) {\it Reduced} $p_T$, base cuts
with $p^b_T > 15$ GeV/c; (dots) {\it Tighter} $\eta$, base cuts with
$|\eta_b| < 1.5$; (dash-dot) {\it Tight} $\eta$ and {\it Loose} $p_T$,
base cuts with the tighter $\eta$ and a looser $p_T$.  {\bf (b)} Plot
of the cross sections relative to the ``Base Cuts'' cross section.}
\end{figure}

In Fig.~\ref{fig:cuts}, we illustrate the effects of these cuts on the
$p\bar{p} \rightarrow h\bar{h} \rightarrow b\bar{b}{\rm Z}^0{\rm Z}^0$
signal.  In Fig.~\ref{fig:cuts}(a), we plot the $p\bar{p} \rightarrow
h\bar{h} \rightarrow b\bar{b}{\rm Z}^0{\rm Z}^0$ cross section for the
following cuts: ``no cuts'' (unconstrained), ``base cuts'' [see
  Table~\ref{tab:tevcuts}], ``tighter pseudorapidity'' cut
($|\eta_b|<1.5$ GeV/c), ``looser transverse momentum'' cut ($p_T>15$
GeV/c), and ``tighter pseudorapidity/looser transverse momentum'' cut
($|\eta_b|<1.5$ and $p_T>15$ GeV/c).  In Fig.~\ref{fig:cuts}(b), we
present each of these cross sections divided by the ``base cuts''
cross section.  Above an $h$ quark mass of $180$ GeV/c$^2$, loosening
the transverse momentum cut from $25$ GeV/c to $15$ GeV/c on each $b$
quark increases the cross section by less than $10$\%.  Tightening the
pseudorapidity cut from $3$ to $1.5$ on {\it both} $b$ quarks reduces
the cross section by roughly 20\%.  Modified constraints on the $b$
quark pseudorapidity may be used to limit the reduction in the cross
section to less than 20\%.  For example, one may require at least one
$b$ quark to be ``tight'' ($\eta_b < 1.5$) and the other $b$ quark to
be ``loose'' ($\eta_b < 3$). 

\begin{figure}
\includegraphics[scale=1.2]{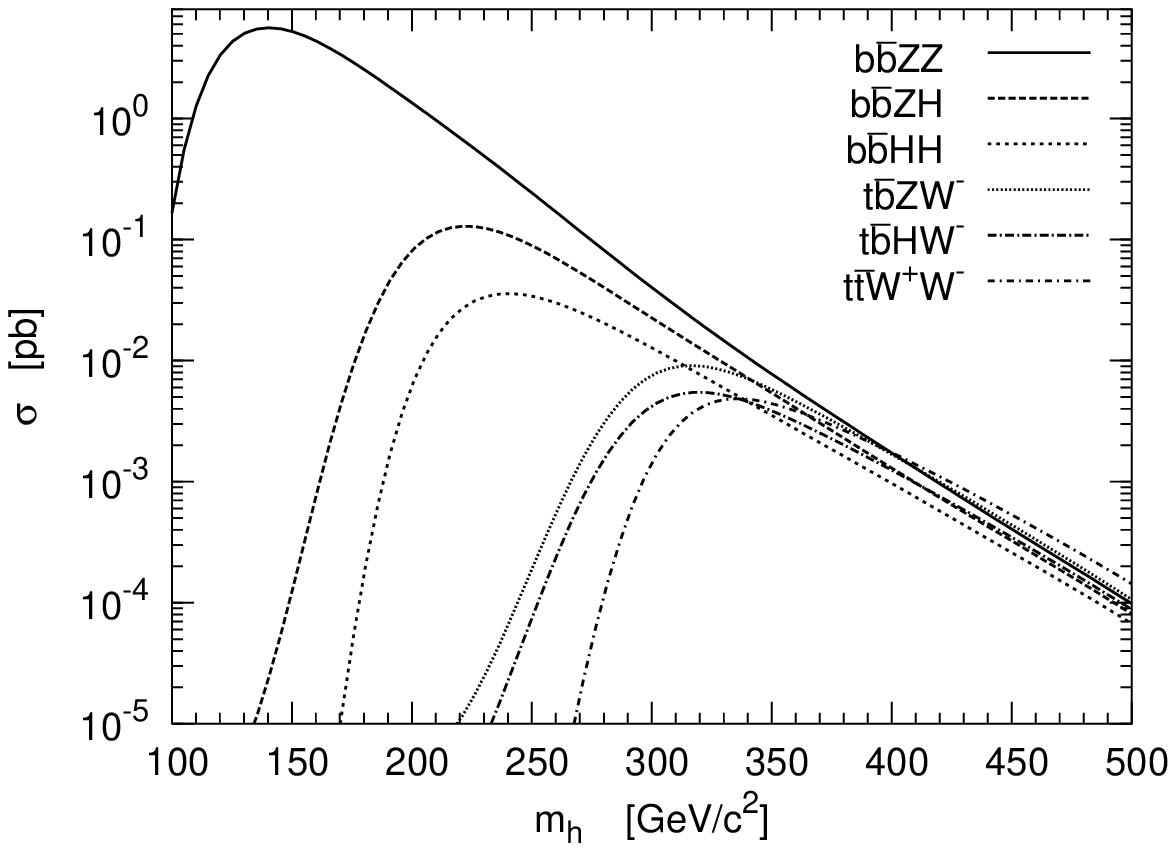}
\caption{\label{fig:xsectev}Cross sections for $h\bar{h}$ production
and primary decay at the Fermilab Tevatron.  In this plot the Higgs
mass is $M_H = 150$ GeV/c$^2$, $\xi = 1$, and $Q=\sqrt{\hat{s}}$.}
\end{figure}

The cross sections for $h\bar{h}$ production and (primary $h$ quark)
decay are shown in Fig.~\ref{fig:xsectev}.  The new mixing parameter
$\xi$ is set to $1$ and the $h$ quark mass, $m_h$, runs from $100$ -- $500$ GeV/c$^2$.
The cross sections for each of the primary decay modes fall rapidly as
the $h$ quark mass is increased.  Because of the transverse momentum cut
on the $b$ quarks, primary decay modes containing at least one $b$
quark exhibit exaggerated rounding of peaks in the cross section.  Cross
sections for modes in which the decay products of the $h$ quark are
more massive than the $h$ quark are suppressed.  In particular, the
cross section for the $t\bar{t}\,{\rm W}^{+}{\rm W}^{-}$ mode is
suppressed until $m_h$ $\ge$ $m_t + M_W$.  As a result of the
suppression and the CM energy of the Tevatron, cross sections for
modes in which at least one $h$ quark decays to a top quark are small
(less than $10$ fb).

At an integrated luminosity of ($1$, $10$) fb$^{-1}$, the largest
accessible $h$ quark mass (at $\xi = 1$ and $M_H=150$ GeV/c$^2$) for any of the
primary $h$ decay modes is $\sim$ ($420$, $500$) GeV/c$^2$ (modes produce one
signal event).  The most promising of these primary decay modes are
the $b\bar{b}\,{\rm Z}^0{\rm Z}^0$ and $b\bar{b}\,{\rm H}^0{\rm Z}^0$
modes.  Each of these modes has a ``large'' cross section below an $h$
quark mass of $300$ GeV/c$^2$, and their respective cross sections are
comparable to the other modes above an $h$ quark mass of $300$
GeV/c$^2$.  As discussed in the signatures section, these modes can
give rise to clean distinctive signatures at the Tevatron.  On the
other hand, modes in which at least one $h$ quark decays to the $t{\rm
  W}^{-}$ channel have relatively small cross sections at the
Tevatron.  Subsequent decay of the top quark often leads to
complicated final states.

\begin{figure}
\begin{tabular}{cc}
  \includegraphics[scale=0.6]{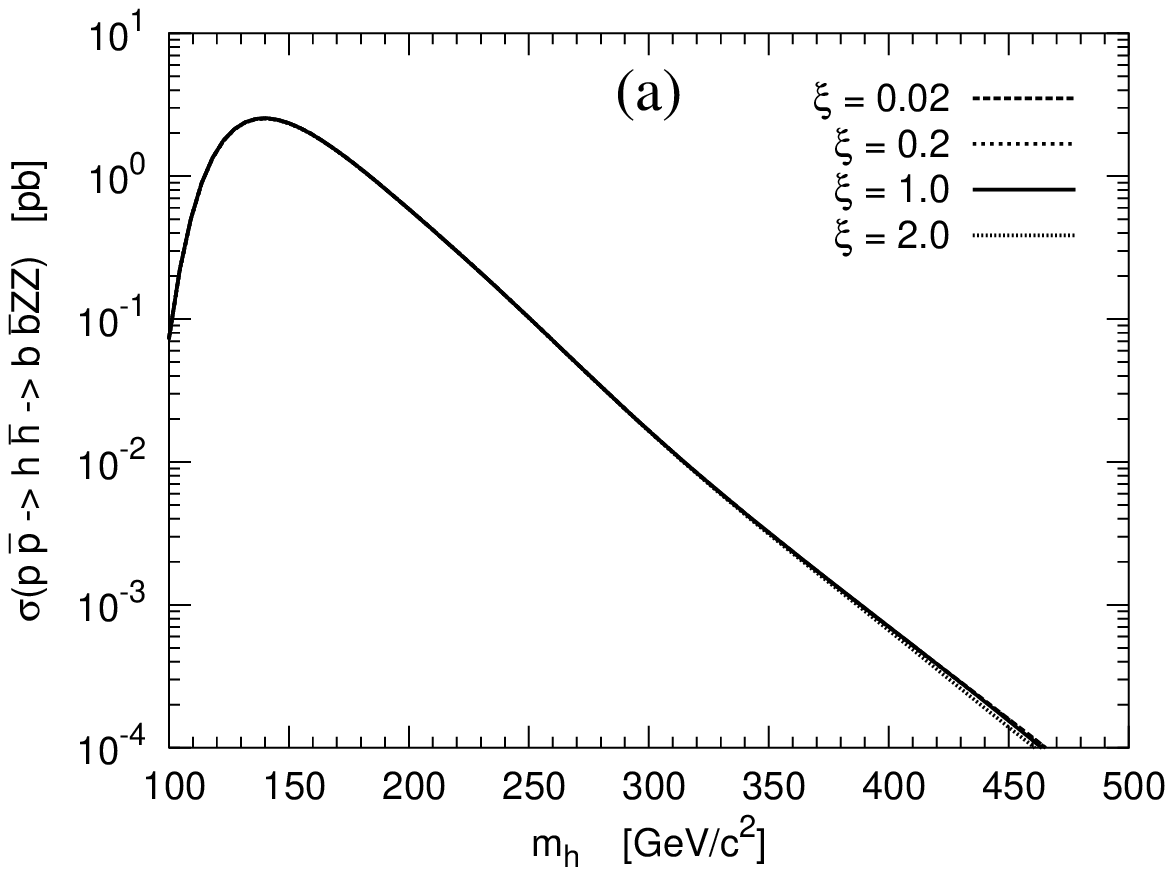} &
  \includegraphics[scale=0.6]{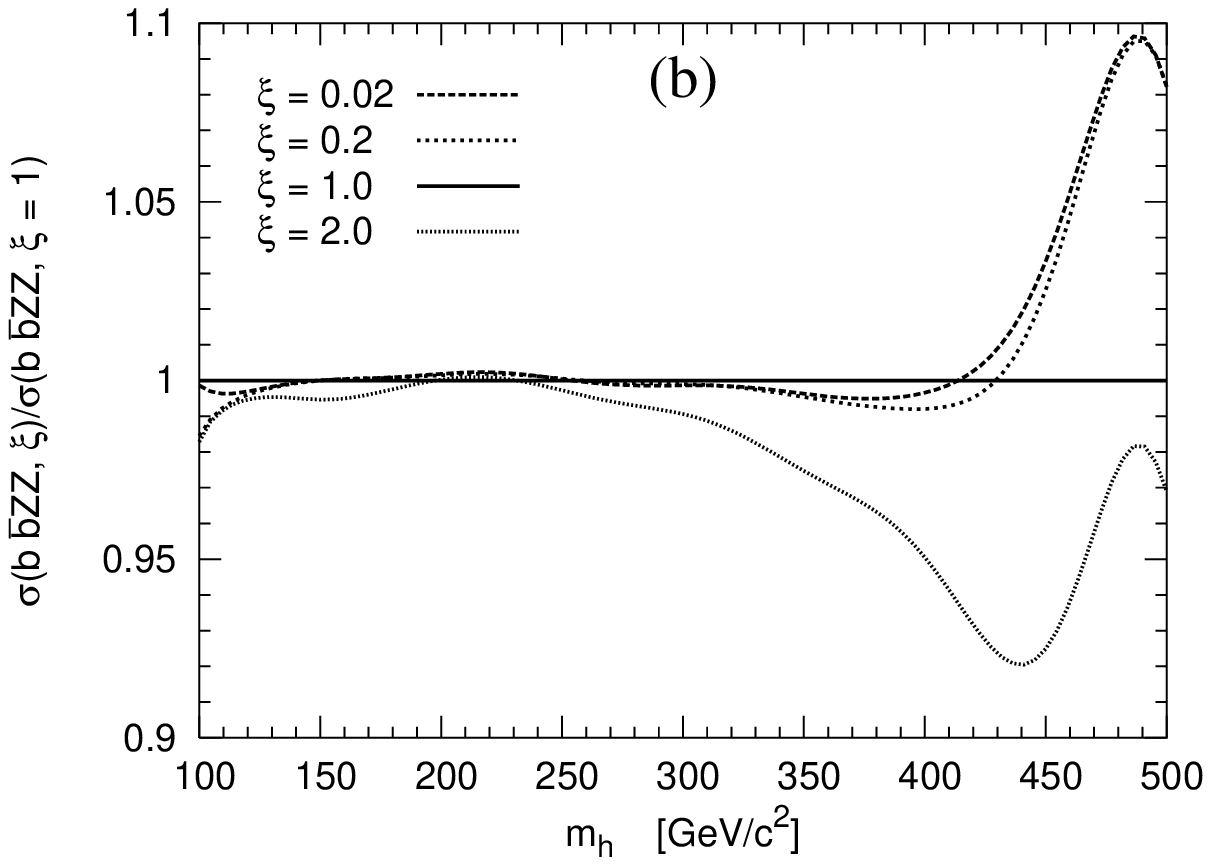} \\
  \includegraphics[scale=0.6]{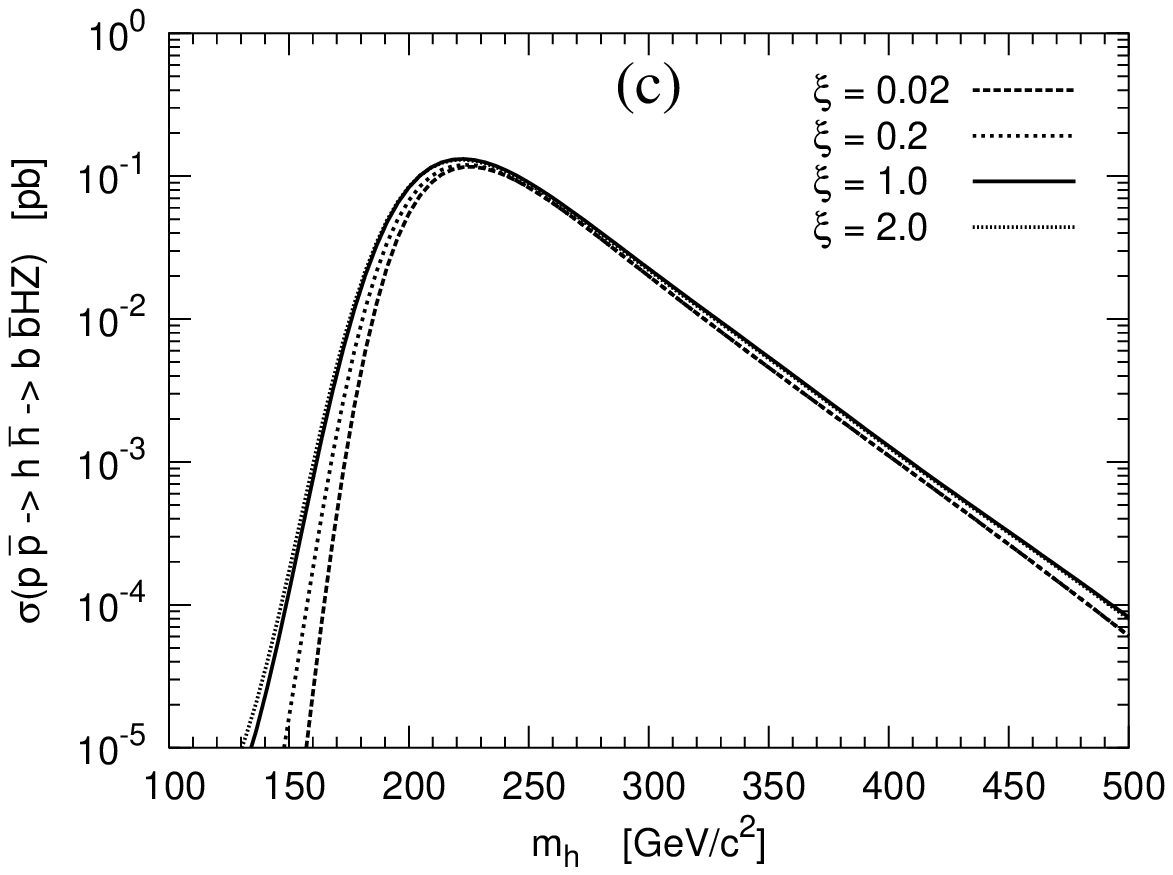} &
  \includegraphics[scale=0.6]{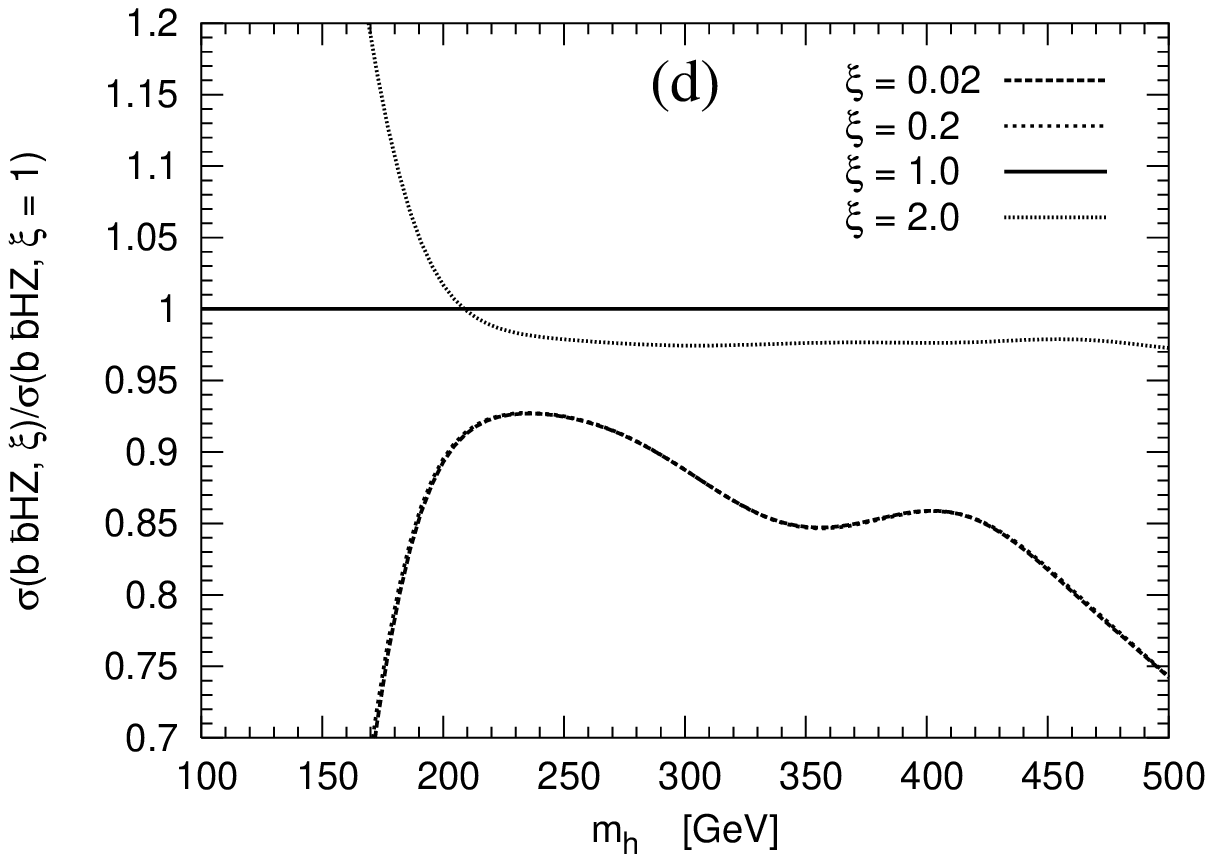} \\
\end{tabular}
\caption{\label{fig:xsectevxi}Effect of the mixing parameter $\xi$ on
the ``primary decay'' cross sections.  The curves in each of these
plots correspond to four different choices of the $\xi$ mixing
parameter ($\xi=0.02$, $0.2$, $1.0$, and $2.0$).  In each of these
plots, the Higgs mass is $150$ GeV/c$^2$. {\bf (a)} Plot of the cross
section for $p\bar{p} \rightarrow h\bar{h} \rightarrow b\bar{b}\,{\rm
Z}^0{\rm Z}^0$ as a function of $h$ quark mass. {\bf (b)} Plot of each
of the $b\bar{b}\,{\rm Z}^0{\rm Z}^0$ cross sections relative to
$\xi=1$ cross section.  {\bf (c)} Plot of the cross section for
$p\bar{p} \rightarrow h\bar{h} \rightarrow b\bar{b}\,{\rm H}^0{\rm
Z}^0$ as a function of $h$ quark mass. {\bf (d)} Plot of each of the
$b\bar{b}\,{\rm H}^0{\rm Z}^0$ cross sections relative to $\xi=1$
cross section.}
\end{figure}

\noindent
{\bf Effect of the $\xi$ Mixing Parameter on Cross Sections:} \\
Before we discuss signatures of $h\bar{h}$ pair production at the
Tevatron, we address the dependence of the cross
sections on the mixing parameter, $\xi$, and on the Higgs mass.  In
Fig.~\ref{fig:xsectevxi}(a) and Fig.~\ref{fig:xsectevxi}(c), we plot
the $p\bar{p} \rightarrow h\bar{h} \rightarrow b\bar{b}{\rm Z}^0{\rm Z}^0$
and $p\bar{p} \rightarrow h\bar{h} \rightarrow b\bar{b}{\rm H}^0{\rm
Z}^0$ cross sections for a Higgs mass of $150$ GeV/c$^2$ and for four
choices of the $\xi$ parameter, $\xi= 0.02$, $0.2$, $1$, $2$~\cite{footnote:reparam}.  In
Fig.~\ref{fig:xsectevxi}(b) and Fig.~\ref{fig:xsectevxi}(d), we
plot the $\xi=0.02$, $0.2$, $1$, $2$ cross sections divided by the
$\xi = 1$ cross section.  In both the $b\bar{b}\,{\rm Z}^0{\rm Z}^0$ and
the $b\bar{b}\,{\rm H}^0{\rm Z}^0$ modes, the cross sections are
weakly dependent on the $\xi$ parameter.  In the $h$ quark mass range
of $100$ -- $500$ GeV/c$^2$, the change in the $\xi$ parameter causes
no more than a 10\% change in the $b\bar{b}\,{\rm Z}^0{\rm Z}^0$ cross
section and no more than a 25\% change in the $b\bar{b}\,{\rm H}^0{\rm
Z}^0$ cross section.  Therefore, reasonable changes in the $\xi$ parameter
should not significantly impact the mass reach capability at the
Fermilab Tevatron.  In the remainder of this paper, we restrict our
analysis to $\xi=1$.

\begin{figure}
\begin{tabular}{c}
  \resizebox{100mm}{!}{\includegraphics{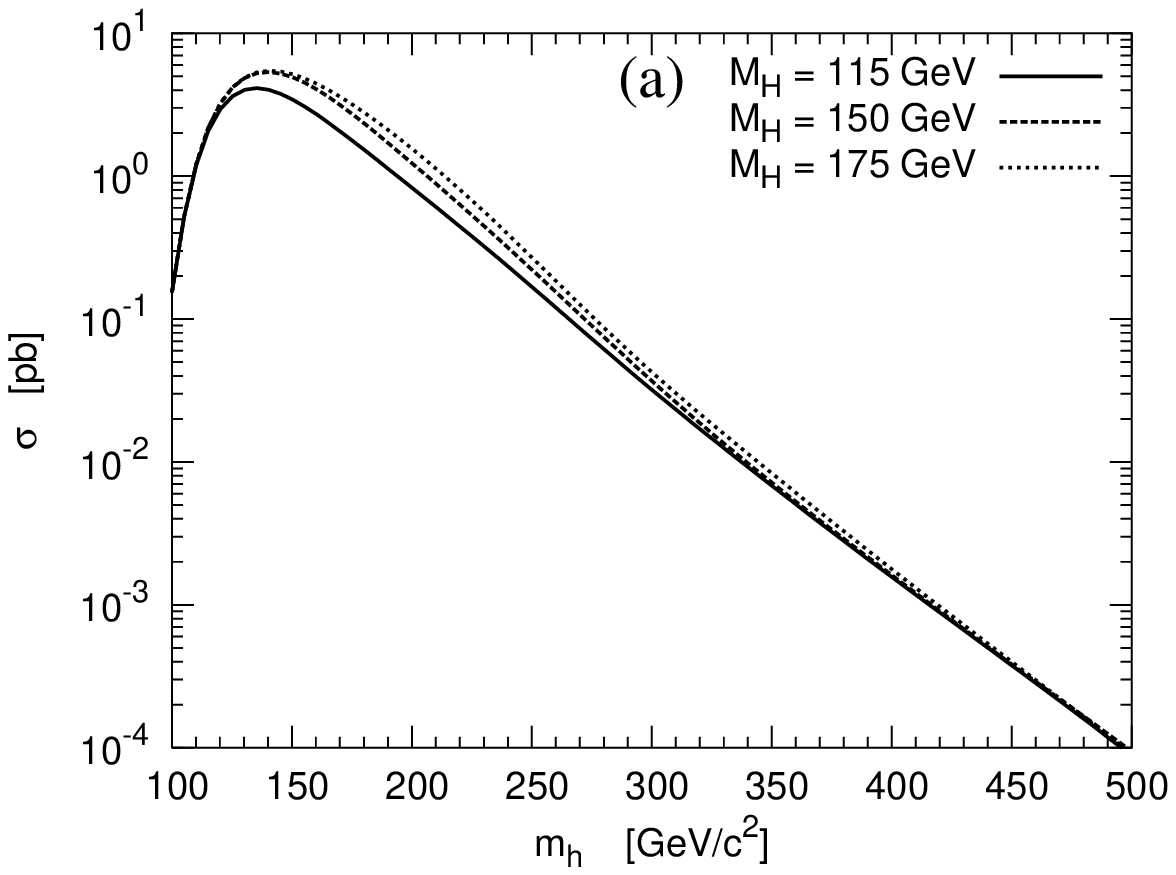}} \\
  \resizebox{100mm}{!}{\includegraphics{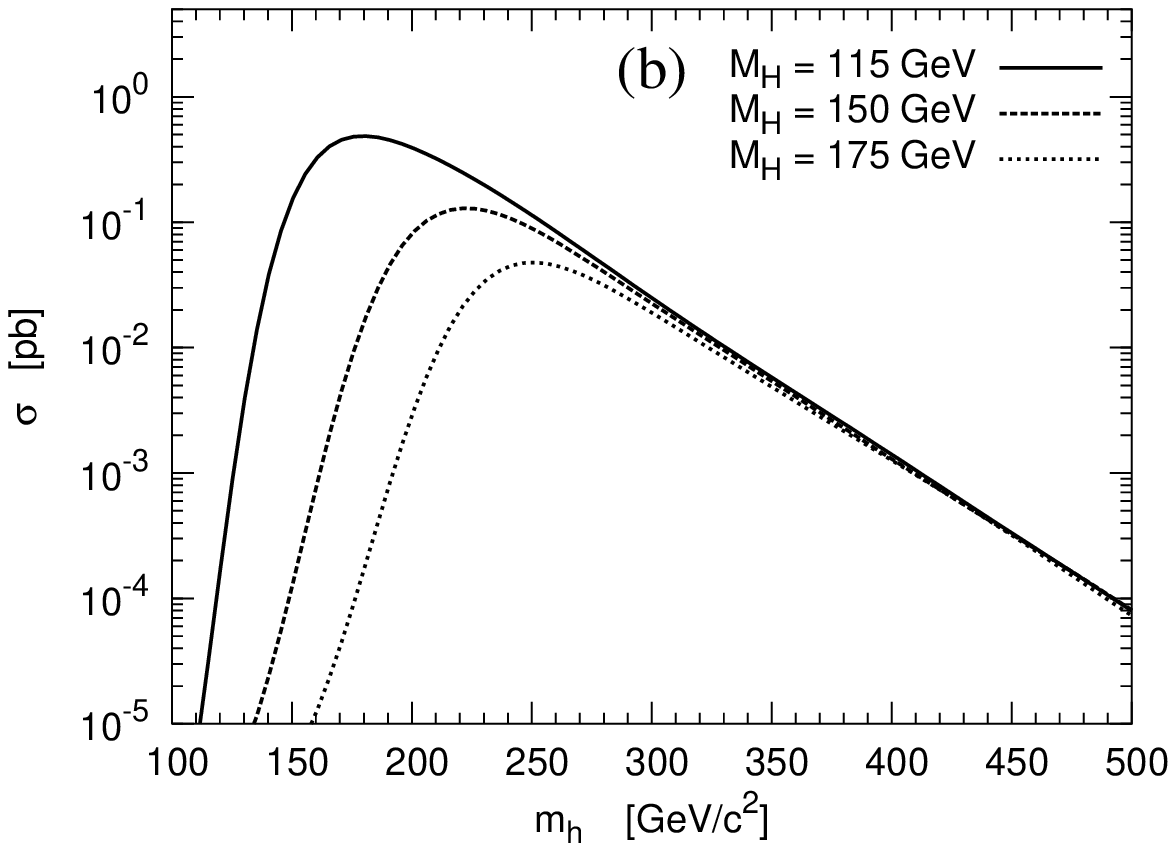}} 
\end{tabular}
\caption{\label{fig:xsectevhiggs}Effect of the Higgs mass, $M_H$, on
the primary decay cross sections.  In each of these plots the mixing
parameter, $\xi$, is fixed at 1, and the curves correspond to
different Higgs masses ($M_H = 115$, $150$, $175$ GeV/c$^2$).  {\bf
(a)} Cross sections for $p\bar{p} \rightarrow h\bar{h} \rightarrow
b\bar{b}\,{\rm Z}^0{\rm Z}^0$ as a function of $h$ quark mass, and
{\bf (b)} Cross section of $p\bar{p} \rightarrow h\bar{h} \rightarrow
b\bar{b}\,{\rm H}^0{\rm Z}^0$ as a function of $h$ quark mass.}
\end{figure}

\noindent
{\bf Effect of Higgs Mass on Cross Sections:} \\ 
In Fig.~\ref{fig:xsectevhiggs}, we plot the $b\bar{b}\,{\rm Z}^0{\rm
Z}^0$ and $b\bar{b}\,{\rm H}^0{\rm Z}^0$ cross sections for three
choices of the Higgs boson mass: $M_H = 115$,
$150$, and $175$ GeV/c$^2$.  At tree level, the $b\bar{b}\,{\rm Z}^0{\rm
Z}^0$ cross section depends on the Higgs mass through the width of the
$h$ quark.  In Fig.~\ref{fig:xsectevhiggs}(a) a heavier Higgs boson
leads to an enhancement in the $b\bar{b}\,{\rm Z}^0{\rm Z}^0$ cross
section.  This can be understood as a suppression of the H$^0\,d_i$
decay channel in the $h$ quark branching ratio [see
  Fig.~\ref{fig:hwid}(b)].

In the $b\bar{b}\,{\rm H}^0{\rm Z}^0$ cross section the Higgs boson is
taken as an ``external particle'' in the Feynman diagrams.  Therefore,
the Higgs boson mass enters the cross section through the phase space
integration and the expression for the $h$ quark width.  In
Fig.~\ref{fig:xsectevhiggs}(b) one observes that by reducing the Higgs
boson mass to $115$ GeV/c$^2$, the cross section is enhanced for the mass
range of $100$ -- $300$ GeV/c$^2$.  On the other hand, if we
increase the mass of the Higgs boson to $175$ GeV, the cross section is
reduced in the same region.  Above an $h$ quark mass of $300$
GeV/c$^2$, changes in the Higgs mass result in small changes to the
$b\bar{b}\,{\rm H}^0{\rm Z}^0$ cross section.  The negative
correlation between the size of the $b\bar{b}\,{\rm H}^0{\rm Z}^0$
cross section and the mass of the Higgs boson below $300$ GeV/c$^2$
follows from the suppression of phase space by the Higgs mass.  In the
remainder of this paper, we consider two Higgs boson masses, $115$
GeV/c$^2$ and $150$ GeV/c$^2$. 

\subsubsection{Signatures}
\label{subsec:signatures}
In this section, we investigate final state signatures for
$h\bar{h}$ pair production at the Tevatron.  To facilitate the
discussion, we categorize signatures based on the primary decay
modes of the $h\bar{h}$ pair.  

Below an $h$ quark mass of $300$ GeV/c$^2$, the dominant primary decay
modes are the $b\bar{b}\,{\rm Z}^0{\rm Z}^0$ and $b\bar{b}\,{\rm H}^0{\rm Z}^0$
modes.  Above an $h$ quark mass of $300$ GeV/c$^2$, each of the
primary decay modes [see Fig.~\ref{fig:xsectev}] are comparable in
size.  Therefore, a number of signatures arising from the decay of the
$b\bar{b}\,{\rm Z}^0{\rm Z}^0$ and the $b\bar{b}\,{\rm H}^0{\rm Z}^0$
modes will be important for an $h\bar{h}$ search.  Though cross sections
from the other decay modes are comparable to the $b\bar{b}\,{\rm
Z}^0{\rm Z}^0$ and the $b\bar{b}\,{\rm H}^0{\rm Z}^0$ modes when the
$h$ quark mass is greater than $300$ GeV/c$^2$, signatures arising
from these modes are often challenging experimentally.

We do not consider the $t\bar{b}\,{\rm Z}^0{\rm W}^-$ ($b\bar{t}\,{\rm
Z}^0{\rm W}^+$), $t\bar{b}\,{\rm H}^0{\rm W}^-$ ($b\bar{t}\,{\rm
H}^0{\rm W}^+$), and $t\bar{t}\,{\rm W}^+{\rm W}^-$ modes because the
decaying $t$ quark produces a $b$ quark and another ${\rm W}^{\pm}$
boson.  As a result, signatures arising from these modes have at least
two more final-state particles than the $b\bar{b}\,{\rm Z}^0{\rm Z}^0$
mode.  At the Tevatron, small cross sections in conjunction with
complicated event topologies, small branching ratios, and detector
effects lead to little or no mass reach for these modes.

For example, signatures arising from the $t\bar{t}\,{\rm W}^+{\rm
W}^-$ mode have complicated final states.  In this
mode, each $t$ quark decays to a $b$ quark and a ${\rm W}^{\pm}$
boson: $(b{\rm W}^+)_{t}(\bar{b}{\rm W}^-)_{t}\,{\rm W}^+{\rm W}^-$.
In our notation, the ``$t$'' subscript in ``$(\cdot)_t$'' indicates that the
quantities enclosed in the parentheses have an invariant mass of the
top quark, ${\rm m}_t$.  Each of the $4$ ${\rm W}^{\pm}$ bosons decay
to either a quark/anti-quark pair ({\it hadronically}) or to a charged
lepton/neutrino pair ({\it leptonically}).  In the detector the
quark/anti-quark pair hadronize into two jets ($jj$), and the charged
lepton/neutrino pair manifest as a lepton~\cite{footnote:notau} and
missing transverse energy ($l^{\pm}\sla{\rm E}_{\rm T}$).  Two possible final state
signatures are: ``$b\bar{b}(jj)_{\rm W}(jj)_{\rm W}(jj)_{\rm
W}(jj)_{\rm W}$'' and ``$b\bar{b} l^- l^+ l^- l^+ \sla{\rm E}_{\rm
T}$''.  The multijet signature has a complicated event topology. The
final state contains two jets associated with $b$ quarks and $8$ jets
arising from hadronization of light quarks.  In order to reconstruct the
$h\bar{h}$ parentage of this signature, one would need excellent dijet
mass resolution to overcome combinatoric challenges.  The
fully-leptonic signature suffers from small branching ratios; in addition,
the presence of multiple uncorrelated neutrinos make it cumbersome
to reconstruct the parentage of the final state leptons.  Other
signatures arising from the $t\bar{t}\,{\rm W}^+{\rm W}^-$ mode suffer
from a combination of the challenges outlined in the {\it full-jets}
and the {\it fully-leptonic} signatures. 

In addition to the charged-current decay modes of the $h\bar{h}$, we
do not consider signatures arising from the $b\bar{b}\,{\rm H}^0{\rm
H}^0$ mode.  Though the 6 $b$ quark signature arising from the decay
of each Higgs boson to $b$ quarks is interesting, the $b\bar{b}\,{\rm
H}^0{\rm Z}^0$ mode has a larger cross section and it leads to cleaner
signatures.  

At the Tevatron, the most promising channels for the discovery of a
down-type isosinglet quark are the $b\bar{b}\,{\rm Z}^0{\rm Z}^0$ and
the $b\bar{b}\,{\rm H}^0{\rm Z}^0$ channels.  These decay channels
arise from the decay of each of the $h$ quarks via a tree-level
flavor-changing neutral interaction (mediated by ${\rm H}^0$ or ${\rm Z}^0$)
to a $b$ quark.  Jets associated with hadronized $b$ quarks in the
final state are powerful objects for analysis thanks to the $b$-jet
identification capabilities at both detector facilities (CDF and D\O\spac).

To obtain a more realistic description of an $h\bar{h}$ event at the
Tevatron, we decay the gauge bosons in the $b\bar{b}\,{\rm Z}^0{\rm
  Z}^0$ and $b\bar{b}\,{\rm H}^0{\rm Z}^0$ channels.  We generate
unweighted events for each of these primary decay modes, and we decay
the gauge bosons to ``final-state'' SM particles (leptons, quarks, and
gluons).  When a ${\rm Z}^0$ or a ${\rm H}^0$ boson is decayed, we
introduce a Gaussian smear on the reconstructed mass of the decay
pair to approximate detector resolution effects.  We take the dijet
mass resolution, $\sigma(M_{jj})/M_{jj}$, of a jet pair to be
10\%~\cite{footnote:smearjj, flow}.  The energy resolution and transverse
momentum resolution from Run I of the CDF detector are used to define
the dilepton mass resolution from electrons and
muons~\cite{Affolder:2000bp}.  Using the {\it signal} events, we
determine the fraction of events that pass a set of detection cuts.
From this fraction and the $b\bar{b}\,{\rm Z}^0{\rm Z}^0$ and
$b\bar{b}\,{\rm H}^0{\rm Z}^0$ cross sections, the cross section of
$h\bar{h}$ events decaying to a particular signature is determined.

In Table~\ref{tab:tevbkgrndcuts} we present the angular and transverse
momentum cuts imposed on the events.  These cuts are applied at the
parton level; the quarks and gluons have not been allowed to
hadronize.  In our notation, ``$l$'' and ``$j$'' refer to a light
charged lepton ($l=$ $e$ or $\mu$) and a jet ($j=$ $u$, $c$, $d$, $s$,
$g$), respectively.  
\begin{table}
\caption{\label{tab:tevbkgrndcuts}Cuts applied to ``final state''
partons.  $l$ refers to either of the ``light'' charged leptons, $e$
or $\mu$.  $j$ refers to the light quarks and gluon -- all of which
would hadronize to form a jet.  Using the ``1'' and ``2'' subscripts,
we distinguish between $b$ quarks arising from the primary decay of
the $h$ quark ($b_1$) and from the subsequent decay of massive gauge
bosons ($b_2$).}
\begin{ruledtabular}
\begin{tabular}{ c c c }
Parameter                 & Minimum Value & Max Value \\ \hline
$\eta_{\rm b}$, $\eta_l$, $\eta_{\rm j}$            &  $-3.0$         &   $3.0$  \\ 
$p_T^{\rm b_1}$                                     &  $25.0$         &    --    \\ 
$p_T^{\, l}$, $p_T^{\, \rm j}$, $p_T^{\rm b_2}$     &  $10.0$         &    --    \\ 
$\Delta$R($i$,\,$k$)                                &   $0.4$         &    --    \\ 
\end{tabular}
\end{ruledtabular}
\end{table}

In addition to the cuts delineated in Table~\ref{tab:tevbkgrndcuts},
dijet and dilepton mass cuts are applied.  In signatures where a ${\rm
Z}^0$ boson decays to either a charged-lepton pair or a jet pair, the
reconstructed mass of that pair should be close to the mass of the ${\rm
Z}^0$ boson.  Therefore, we require the mass of a dijet/dilepton pair
be within $\pm (0.1\times M_Z)$.  Complications may arise when more
than two charged leptons or two jets are in the final state.  In
such instances, we define ``Z$^0$-like'' lepton or jet pairs as those with
mass closest to the ${\rm Z}^0$ mass.  In analogy to the ${\rm
Z}^0$ case, we impose a dijet reconstruction cut on the $b$ jet decay
products of the ${\rm H}^0$ boson.  Also, if a Higgs boson decays to a
$b\bar{b}$ pair, there will be at least $4$ $b$ jets in the final state --
two $b$ quarks from the decay of the $h$ quarks and two $b$ quarks
from the Higgs decay.  Rather than requiring each $b$ jet to have a transverse
momentum greater than $25$ GeV/c, we require that at least two {\it
  non} ``H$^0$-like'' $b$ jets have a transverse momentum greater than
$25$ GeV/c.  The remaining $b$ jets must have a transverse momentum
greater than $10$ GeV/c.  

We also require additional cuts on the reconstructed $h$ quark mass.
For both the $b\bar{b}\,{\rm Z}^0{\rm Z}^0$ and the $b\bar{b}\,{\rm
H}^0{\rm Z}^0$ modes, the $h$ quark mass can be reconstructed from
the appropriate $b$ jet and gauge boson combinations.  We constrain
the invariant mass of a $b$ quark and one of the Z$^0$-like or
H$^0$-like final-state particle pairs be equal (within resolution) to
the invariant mass of the other $b$ quark and gauge boson decay
products.  In addition, each of these ``$h$ quark legs'' must equal
(within resolution) the desired $h$ quark mass ({\it e.g.} $200$,
$250$, or $300$ GeV/c$^2$).  If one of the gauge bosons cannot be reconstructed
({\it e.g.} ${\rm Z}^0$ decays to neutrinos), then the invariant mass
of one of the $b$ quarks and the other {\it reconstructible} ${\rm
  Z}^0$-like decay products must be equal (within resolution) to the
$h$ quark mass.

With a set of formal cuts in place, we investigate specific
final-state signatures arising from the $b\bar{b}\,{\rm Z}^0{\rm Z}^0$ and
$b\bar{b}\,{\rm H}^0{\rm Z}^0$ channels.  In Table~\ref{tab:signal} we
list five signatures for the $b\bar{b}\,{\rm Z}^0{\rm Z}^0$ mode and
three signatures for the $b\bar{b}\,{\rm H}^0{\rm Z}^0$ mode.  We
consider signatures arising from ${\rm Z}^0$ decays to a jet pair
($jj$), a $b$-jet pair ($bb$), a neutrino pair (missing transverse
energy, $\sla{\rm E}_{\rm T}$), and a pair of ``light'' charged
leptons ($l=$ $e$, $\mu$).  Signatures resulting from the decay of the
Higgs boson to a $b$-jet pair are also considered.  The cross sections
contained in Table~\ref{tab:signal} do not account for $b$-tagging
efficiencies [$\epsilon_b = 100\%$].  In
Section~\ref{subsec:backgrounds} we loosen this constraint to
determine estimates of the $h$ quark mass reach at the Tevatron. 
\begin{table}
\caption{\label{tab:signal}The signal cross section (in fb) for
$h\bar{h}$ pair production and subsequent decay into $7$ final-state
signatures.  We use the `$(\cdot)_{\rm Z}$' notation to indicate that
the quantities enclosed in the parentheses should have an invariant
mass equal to the ${\rm Z}^0$ boson mass.  The cross sections are
presented for three choices of the $h$ quark mass ($m_h = 200$, $250$,
and $300$ GeV/c$^2$) and for two choices of the Higgs mass ($M_H =
115$ and $150$ GeV/c$^2$).  These cross sections assume perfect
$b$-jet tagging efficiencies, $\epsilon_b = 100\%$.}
\begin{ruledtabular}
\begin{tabular}{l | c c c | c c c}
Signature                                  & \multicolumn{3}{c|}{$\sigma$ (fb)}              & \multicolumn{3}{c}{$\sigma$ (fb)} \\
$m_h$ (GeV/c$^2$)                          & 200                    & 250                    & 300  
                                           & 200                    & 250                    & 300                    \\
$M_H$ (GeV/c$^2$)                          & 115                    & 115                    & 115  
                                           & 150                    & 150                    & 150                    \\ \hline
$bb(jj)_{\rm Z}(jj)_{\rm Z}$               &  50                    &  13                    &   1.9                 
                                           &  79                    &  18                    &   0.90                 \\
$bb(jj)_{\rm Z}(l^+l^-)_{\rm Z}$           &  19                    &   4.9                  &   0.72                 
                                           &  28                    &   6.4                  &   0.30                 \\
$bb(jj)_{\rm Z}{\sla {\rm E}_{\rm T}}$     &  90                    &  23                    &   3.2                 
                                           & 140                    &  32                    &   1.5                  \\
$bb(l^+l^-)_{\rm Z}{\sla {\rm E}_{\rm T}}$ &  15                    &   4.0                  &   0.55                 
                                           &  26                    &   5.6                  &   0.29                 \\ 
$bb(bb)_{\rm Z}{\sla {\rm E}_{\rm T}}$     &  29                    &   7.3                  &   1.0                     
                                           &  44                    &   9.4                  &   0.45                 \\ \hline
$bb(bb)_{\rm H}(jj)_{\rm Z}$               &  46                    &  15                    &   2.5                 
                                           &   4.9                  &   3.4                  &   0.43                 \\ 
$bb(bb)_{\rm H}(l^+l^-)_{\rm Z}$           &   7.9                  &   2.9                  &   0.45                 
                                           &   0.74                 &   0.59                 &   0.080                \\ 
$bb(bb)_{\rm H}{\sla {\rm E}_{\rm T}}$     &  37                    &  12                    &   1.9                 
                                           &   4.0                  &  2.3                   &   0.36                 \\
\end{tabular}
\end{ruledtabular}
\end{table}

As seen in Table~\ref{tab:signal}, the most promising decay
signatures in the $b\bar{b}\,{\rm Z}^0{\rm Z}^0$ channel are
$b\bar{b}(jj)_Z(jj)_Z$, $b\bar{b}(jj)_Z(\sla{\rm E}_{\rm T})_Z$, and
$b\bar{b}(jj)_Z(l^+ l^-)_Z$.  These decay signatures benefit from at
least one of the ${\rm Z}^0$ bosons decaying hadronically.  The mode
in which both ${\rm Z}^0$ bosons decay to light quarks (full-jets
mode) appears to provide the best reach in $h$ quark mass.  The
$b\bar{b}(jj)_Z(\sla {\rm E}_{\rm T})_Z$ and $b\bar{b}(jj)_Z(l^+
l^-)_Z$ signatures are relatively clean with slightly diminished
signal.  If the $h$ quark mass is $250$ GeV/c$^2$ and the mass of the
Higgs boson is $150$ GeV/c$^2$, one expects to produce ($18$, $180$)
$bb(jj)_{\rm Z}(jj)_{\rm Z}$ events, ($6.4$, $64$) $bb(jj)_{\rm
  Z}(l^+l^-)_{\rm Z}$ events, and ($32$, $320$) $bb(jj)_{\rm
  Z}{\sla{\rm E}_{\rm T}}$ events in ($1$, $10$) fb$^{-1}$ of data at
the Tevatron.  Many of the $b\bar{b}\,{\rm Z}^0{\rm Z}^0$ signatures
are interesting and are similar to the signatures of $t\bar{t}$
production at the Tevatron. In Section~\ref{subsec:backgrounds} we
find that some of these signatures are closely related to the
$t\bar{t}$ production signal, resulting in large backgrounds.

Next, we consider final state signatures arising from the
$b\bar{b}\,{\rm H}^0{\rm Z}^0$ channel.  In this mode the Higgs
boson is produced via the tree-level FCNC decay of the $h$ quark.
Since the strength of the coupling to the $h$ quark is proportional to
the product of the flavor-changing interaction and the mass of the $h$ quark [see
Fig.~\ref{fig:feynman}], the Higgs coupling to the $h$ quark may be
sizable.  This leads to an intriguing scenario in which the discovery
of an isosinglet quark may lead to the discovery of the Higgs boson.
We consider signatures arising from the decay of the Higgs boson to
$b$ quarks.  Decays of the Higgs boson to $W^{\pm}$ pairs lead to
complicated signatures that are not ideal for an $h$ search at the Tevatron.

In Table~\ref{tab:signal}, we consider three signatures arising
from the $b\bar{b}\,{\rm H}^0{\rm Z}^0$ channel.  The ${\rm Z}^0$
boson decays to either a quark/anti-quark pair, a pair of charged
leptons, or a pair of neutrinos.  As discussed above, jets in the
final state result from the hadronization of the light quarks, and
missing transverse energy ($\sla {\rm  E}_{\rm T}$) results from the
decay of a ${\rm Z}^0$ boson to neutrinos.

As expected, the branching ratio of the Higgs boson to a $b$ quark pair
is significantly reduced as one increases the Higgs boson mass from
$115$ GeV/c$^2$ to $150$ GeV/c$^2$.  Signatures arising from the
$b\bar{b}\,{\rm H}^0{\rm Z}^0$ channel have the most reach when there
is a {\it light} Higgs boson.  In addition, the $bb(bb)_{\rm
H}(jj)_{\rm Z}$ and the $bb(bb)_{\rm H}{\sla {\rm E}_{\rm T}}$
signatures have much larger cross sections than the $bb(bb)_{\rm
H}(l^+l^-)_{\rm Z}$ signature.  If the $h$ quark mass is $250$
GeV/c$^2$ and the Higgs mass is $115$ GeV/c$^2$, one expects to
produce ($15$, $150$) $bb(bb)_{\rm  H}(jj)_{\rm Z}$ events and ($12$,
$120$) $bb(bb)_{\rm H}{\sla{\rm E}_{\rm T}}$ in ($1$, $10$) fb$^{-1}$
of data at the Fermilab Tevatron. 

\subsubsection{Backgrounds}
\label{subsec:backgrounds}
In the previous section we presented a number of $h\bar{h}$ signatures
and their expected tree-level cross sections at the Tevatron.  To ascertain the
$h$ quark mass reach at the Tevatron, we need to understand the
backgrounds associated with these signatures [see Table~\ref{tab:signal}].

We use the software package MadEvent~\cite{Maltoni:2002qb} to study the
SM backgrounds for these signatures.  MadEvent generates and calculates
the {\it tree-level} contributions to a given process (parton-level calculation)
using helicity amplitude methods~\cite{Murayama:1992gi}.  We do not
consider one-loop contributions to the background.  Though FCNCs occur
at one-loop in the SM, we expect their contribution to the background
to be small.  If one includes one-loop processes in the SM background
calculation, the one-loop contribution to the ISVLQ model should be
included in the signal calculation.

At the parton level, each of the $h\bar{h}$ signatures contain six final
state particles (quarks and leptons).  Though MadEvent can generate the
diagrams and associated matrix element for a signature containing six
``final-state'' particles, the evaluation of some matrix
elements is computationally intractable.  In particular, the
quantum chromodynamic backgrounds to the $bb(jj)_{\rm Z}(jj)_{\rm
Z}$ and the $bb(bb)_{\rm H}(jj)_{\rm Z}$ signatures consist of an
enormous set of tree-level diagrams that overwhelm most computing
clusters.  Therefore, when computationally feasible, we use MadEvent
to calculate the SM background to the $h\bar{h}$ signatures.  The
order of the background calculation will accompany all estimates of
the background cross sections.  For signatures like the $bb(jj)_{\rm
  Z}(jj)_{\rm Z}$, in which the matrix element for the backgrounds are
not calculable using current technology, one would need to use
approximate methods or to measure the background from the Tevatron
data itself.

For each signature, we require background processes to pass the
cuts outlined in Table~\ref{tab:tevbkgrndcuts} and the invariant mass
cuts discussed in Section~\ref{subsec:signatures}.  As with the signal
events, we assume that at the Tevatron the dijet mass resolution can
be approximated by a Gaussian distribution with a resolution of 10\%
of the invariant mass.  The dijet and dilepton cuts and the cuts in
Table~\ref{tab:tevbkgrndcuts} are designed to reduce the size of the background.
For example, the hard cut on the transverse momentum of $b$ quarks
from the primary $h$ quark decay ($p^b_T > 25$ GeV/c$^2$) in
conjunction with the jet separation cut ($\Delta R > 0.4$) helps
reduce backgrounds from gluon splitting to $b\bar{b}$.  The dijet mass
cuts help reduce QCD backgrounds that duplicate our $h\bar{h}$
production signatures.  

In Table~\ref{tab:bkgrnd}, we present the results of our SM background
calculation for each of the signatures listed in
Table~\ref{tab:signal}.  For the SM background calculations we assume
a $b$-jet tagging efficiency of 50\%; if a $b$ jet is {\it not}
tagged, then it is treated as a jet ($j$).  For signatures containing $2$
or $4$ $b$ jets, we require at least $1$ or $3$ $b$-jet tags,
respectively.  As discussed above, we are unable to calculate the
matrix element of the backgrounds for some signatures because of computational
limitations.  In columns four and five of Table~\ref{tab:bkgrnd}, we
indicate the electroweak order of the background for which we were
able and unable to calculate.  In these columns, a number indicates the
number of electroweak vertices in the background calculation, the
``FT'' indicates the calculation is a full tree-level calculation, and
the ``*'' indicates that the background is forced to produce massive
gauge bosons before decaying to the indicated signature.  For example,
we calculated the $bb(jj)_{\rm Z}(l^+l^-)_{\rm Z}$ background
originating from diagrams with two electroweak vertices and with four
electroweak vertices. The background with two electroweak vertices
includes diagrams where the jets do not come from ${\rm Z}^0$ decay.
The background with four electroweak vertices includes all
$b\bar{b} (jj)_Z (l^+l^-)_{\rm Z}$ signatures that arise from
$b\bar{b}\mathcal{XX}$, where ${\mathcal X}$ is a massive gauge
boson, ${\mathcal X} = {\rm Z}^0$, ${\rm W}^{\pm}$, or ${\rm H}^0$.  
\begin{table}
\caption{\label{tab:bkgrnd}Standard Model background to the $h\bar{h}$
pair production signatures outlined in
Section~\ref{subsec:signatures}.  The SM background cross sections are
presented for a Higgs mass of $115$ and of $150$ GeV/c$^2$.  A $b$-jet
tagging efficiency of $50\%$ is used and at least $1$ or $3$ $b$ tags
are required for signatures containing $2$ or $4$ $b$ jets,
respectively.  In the ``EW Order'' columns we indicate the order of
each of the electroweak calculations.  A number in either of these
columns indicates the number of electroweak vertices in the background
calculation, the ``FT'' indicates the calculation is a full tree-level
calculation, and the ``*'' indicates the background is forced to
produce massive gauge bosons before decaying to the indicated signature.} 
\begin{ruledtabular}
\begin{tabular}{l | c c c | c c c | c c}
Signature                                  &
                                           \multicolumn{3}{c|}{$\sigma_{bkgrnd}$(fb)} & \multicolumn{3}{c|}{$\sigma_{bkgrnd}$(fb)} & \multicolumn{2}{c}{EW Order} \\ 
$M_h$ (GeV/c$^2$)                          & 200     & 250     & 300      & 200     & 250     & 300      & (calc)    &    (uncalc)  \\
$M_H$ (GeV/c$^2$)                          & 115     & 115     & 115      & 150     & 150     & 150      &           &              \\ \hline
$bb(jj)_{\rm Z}(jj)_{\rm Z}$               & 110     & 56      &  16      & 110     & 57      & 15       & 2,4$^*$   & $0,>4$       \\
$bb(jj)_{\rm Z}(l^+l^-)_{\rm Z}$           &  0.023  &  0.0098 &  0.0044  &  0.024  &  0.010  &  0.0043  & 2,4$^*$   & $>4$         \\
$bb(jj)_{\rm Z}{\sla {\rm E}_{\rm T}}$     &  1.8    &  0.57   &  0.30    &  1.3    &  0.57   &  0.53    & 2,4$^*$   & $>4$         \\
$bb(l^+l^-)_{\rm Z}{\sla {\rm E}_{\rm T}}$ &  8.4    &  6.1    &  2.6     &  8.4    &  6.1    &  2.6     & FT        &              \\ 
$bb(bb)_{\rm Z}{\sla {\rm E}_{\rm T}}$     &  0.023  &  0.014  &  0.0081  &  0.022  &  0.014  &  0.0079  & 2,4$^*$   & $>4$         \\ \hline
$bb(bb)_{\rm H}(jj)_{\rm Z}$               &  0.059  &  0.0023 &  0.0010  &  0.0025 &  0.0019 &  0.00091 & 2,4$^*$   & $0,>4$       \\ 
$bb(bb)_{\rm H}(l^+l^-)_{\rm Z}$           &  0.0035 &  0.0013 &  0.00047 &  0.0026 &  0.0015 &  0.00047 & 2,4$^*$   & $>4$         \\ 
$bb(bb)_{\rm H}{\sla {\rm E}_{\rm T}}$     &  0.021  &  0.014  &  0.0087  &  0.013  &  0.013  &  0.0090  & 2,4$^*$   & $>4$         \\
\end{tabular}
\end{ruledtabular}
\end{table}

For most of the signatures discussed in Section~\ref{subsec:signatures}, the
SM background appears to be manageable when compared to the signal
cross sections [see Table~\ref{tab:signal}].  The notable exceptions
are the $bb(jj)_{\rm Z}(jj)_{\rm Z}$ and the $bb(l^+l^-)_{\rm Z}{\sla
  {\rm E}_{\rm T}}$ signatures.  The large backgrounds for these
signatures (relative to the other signatures) can be traced to top
quark pair production.  At the Tevatron, $t\bar{t}$ production, like
the $h\bar{h}$ production, proceeds via quark/anti-quark annihilation
and gluon-gluon fusion.  Once the $t\bar{t}$ are produced, each top
quark decays to a $b$ and a ${\rm W}^+$.  Subsequent decay of the
${\rm W}^{\pm}$ bosons results in the following three signatures: ``$b\bar{b}(jj)_{\rm W}(jj)_{\rm W}$'' (full-jets),
``$b\bar{b} (jj)_{\rm W} \, l^{\pm} \sla{\rm E}_{\rm T}$''
(semi-leptonic), and ``$b\bar{b}\, l^{+} l^{-} \sla{\rm E}_{\rm T}$''
(fully leptonic).  We use the `$(\cdot)_{\rm W}$' notation
to indicate that the quantities enclosed in the parentheses have an
invariant mass equal to the ${\rm W}^{\pm}$ boson mass.  

If the dijet mass resolution at the Tevatron were {\it perfect}, then a
jet pair from a ${\rm W}^{\pm}$ decay and a jet pair from a ${\rm
  Z}^0$ decay would always be distinguishable.  However, the dijet mass
resolution at the Tevatron is {\it not} perfect; therefore, a fraction
of the $t\bar{t}$ events decaying to the full-jets mode will mimic the
$bb(jj)_{\rm Z}(jj)_{\rm Z}$ signature.  We find that for a Higgs mass
of $150$ GeV/c$^2$ and an $h$ quark mass of $250$ GeV/c$^2$, the
background to the signature $bb(jj)_{\rm Z}(jj)_{\rm Z}$ is $57$ fb.
The purely QCD component of this background (denoted ``0'' in
Table~\ref{tab:bkgrnd}) was not
calculated.  Though one expects the invariant mass cuts to
substantially reduce this background, it is unlikely that it will be
negligible.  Thus we conclude that the $bb(jj)_{\rm Z}(jj)_{\rm Z}$
signature is background limited and that other $h\bar{h}$ signatures
will provide a better $h$ quark mass reach.  However, if nature
contains an $h$ quark with a mass less than $\sim 250$ GeV/c$^2$, this
signature will provide a channel to measure the mass of the $h$ quark
($h$ quark mass peak).

The $bb(l^+l^-)_{\rm Z}{\sla{\rm E}_{\rm T}}$ signature also has a sizable $t\bar{t}$
background component.  When a $t\bar{t}$ event decays to the fully
leptonic mode, the invariant mass of the two charged leptons {\it may}
be close to the mass of the ${\rm Z}^0$ boson.  One concludes that,
like the $bb(jj)_{\rm Z}(jj)_{\rm Z}$ signature, the $h$ quark mass
reach of the $bb(l^+l^-)_{\rm Z}{\sla{\rm E}_{\rm T}}$ signature is
diminished because of the background.

The $b\bar{b}(jj)_Z(l^+ l^-)_Z$ and $b\bar{b}(jj)_Z \sla{\rm E}_{\rm T}$
signatures provide clean alternatives to the full jets signature.
Unlike the $bb(jj)_{\rm Z}(jj)_{\rm Z}$ signature, the
$b\bar{b}(jj)_Z\sla{\rm E}_{\rm T}$ signature is not afflicted by a large
$t\bar{t}$ background.  The semi-leptonic decay of a $t\bar{t}$ event
can duplicate the $b\bar{b}(jj)_Z\sla{\rm E}_{\rm T}$ signature if the
dijet mass is close to the ${\rm Z}^0$ mass and the charged lepton is
undetected.  In Table~\ref{tab:bkgrnd}, we include this and other
backgrounds to the $b\bar{b}(jj)_Z\sla{\rm E}_{\rm T}$ signature in
which a charged lepton from the decay of a ${\rm W}^{\pm}$ is
undetected.  In order to determine the ``undetected lepton''
background, we assume that this background originates from events in
which the charged lepton travels through an uninstrumented region of the
detector~\cite{footnote:uninstrumented}.  The ``lost-lepton''
background is also included in the $b\bar{b}(l^+l^-)_Z\sla{\rm E}_{\rm
  T}$, $b\bar{b}(bb)_Z\sla{\rm E}_{\rm T}$, and
$b\bar{b}(bb)_H\sla{\rm E}_{\rm T}$ signatures.  Though other
experimental issues, like jet energy mismeasurement of QCD jets, are
likely to increase the background, we do not include these in our
calculation.  

The $b\bar{b}(jj)_Z(l^+ l^-)_Z$ signature also avoids large $t\bar{t}$
backgrounds.  The semi-leptonic decay of a $t\bar{t}$ event can mimic
this signature if the dijet mass is ${\rm Z}^0$-like and the detector
spuriously identifies an additional lepton.  The dilepton mass of the
real and spurious charged leptons must be ${\rm Z}^0$-like.  This
component of the background is expected to be small; therefore, we do
not include it in the background estimate. 

Since the $b\bar{b}(jj)_Z(l^+ l^-)_Z$ and the $b\bar{b}(jj)_Z \sla{\rm
E}_{\rm T}$ are clean signatures with small backgrounds, we expect these modes to
provide the greatest reach at the Tevatron.  In ($1$, $10$) fb$^{-1}$ of data,
the highest $h$ quark mass accessible by the $b\bar{b}(jj)_Z(l^+
l^-)_Z$ and the $b\bar{b}(jj)_Z \sla{\rm E}_{\rm  T}$ signatures are
($230$, $290$) GeV/c$^2$ and ($270$, $320$) GeV/c$^2$, respectively.

While the $b\bar{b}\,{\rm Z}^0{\rm Z}^0$ mode is likely to provide the
best reach for an $h$ quark search, the $b\bar{b}\,{\rm H}^0{\rm Z}^0$
mode provides an opportunity to discover the Higgs boson in
conjunction with the $h$ quark.  If nature favors a light Higgs boson
({\it e.g.} $M_H=115$ GeV/c$^2$), the dominant branching ratio of the
Higgs boson is the $b\bar{b}$ mode, $BR({\rm H}^0 \rightarrow
b\bar{b}) = 73.2$\%.  However, if the mass of the Higgs boson is
larger than current electroweak best fits~\cite{mhiggs}, the $b\bar{b}{\rm
H}^0{\rm Z}^0$ mode is less powerful.  For example, at a Higgs mass of
$150$ GeV/c$^2$, the Higgs branching ratio to $b\bar{b}$ is $17.6$\%.

Unlike many of the $b\bar{b}\,{\rm Z}^0{\rm Z}^0$ signatures, the
$b\bar{b}\,{\rm H}^0{\rm Z}^0$ signatures do {\it not} suffer from the
large $t\bar{t}$ background.  The reduction in the $t\bar{t}$
background is primarily because of the $b\bar{b}$ signature from the Higgs
decay (${\rm W}^{\pm}$ can not decay to a $b\bar{b}$ pair).
Though these signatures have low SM backgrounds, $b$-tagging
efficiencies will reduce the expected signal cross section.

Standard Model background to each of the three $b\bar{b}\,{\rm
H}^0{\rm Z}^0$ signatures is small [see Table~\ref{tab:bkgrnd}].
Because of computational limitations, the pure QCD component of the
$b\bar{b}(bb)_H(jj)_Z$ background  (denoted ``0'' in
Table~\ref{tab:bkgrnd}) was not calculated.  As with the background
for the $b\bar{b}(jj)_Z(jj)_Z$ signature, the QCD component of the
$b\bar{b}(bb)_H(jj)_Z$ signature will need to be measured from the
data.  It is unlikely that the background from these processes will be
large.  Assuming a $b$-tagging efficiency of $50$\% and requiring that
three of the four $b$-jets is tagged, for a Higgs mass of $115$
GeV/c$^2$ one expects the mass reach in the
$b\bar{b}(bb)_H\sla{{\rm E}}_{\rm T}$ to be ($230$, $290$) GeV/c$^2$ in
($1$, $10$) fb$^{-1}$ of data.  Moreover, because the background to
the $b\bar{b}(jj)_H(jj)_Z$ signature is not expected to be large, we
expect for a Higgs mass of $115$ GeV/c$^2$ the $h$ mass reach of this
mode to be $\sim$ ($220$, $270$) GeV/c$^2$.

\section{$h\bar{h}$ Production and Decay at the LHC}
\label{sec:panddLHC}
At the CM energy of the CERN LHC ($\sqrt{s} = 14$ TeV), $h\bar{h}$
pair production is dominated by the gluon-gluon fusion subprocess.
Contributions from subprocesses in which a valence quark from one
proton and its anti-particle from the other proton (sea quark)
annihilate via the strong interactions are important but sub-dominate
to the gluon-gluon fusion.  

For the LHC we consider the same primary decay channels of the
$h\bar{h}$ pair as at the Tevatron: $t\bar{t}\,{\rm W}^+{\rm W}^-$,
$b\bar{b}\,{\rm Z}^0{\rm Z}^0$, $b\bar{b}\,{\rm H}^0{\rm H}^0$,
$t\bar{b}\,{\rm W}^-{\rm Z}^0$, $b\bar{b}\,{\rm H}^0{\rm Z}^0$, and
$t\bar{b}\,{\rm W}^-{\rm H}^0$.  We also impose the same set of loose
cuts on the primary decay products of the $h$ quark [see
Table~\ref{tab:tevcuts}].  This ensures that $b$ quarks produced from
the primary decay of the $h\bar{h}$ pair conform to {\it basic}
geometry and event selection requirements of the detectors.

\begin{figure}
\includegraphics[scale=1.2]{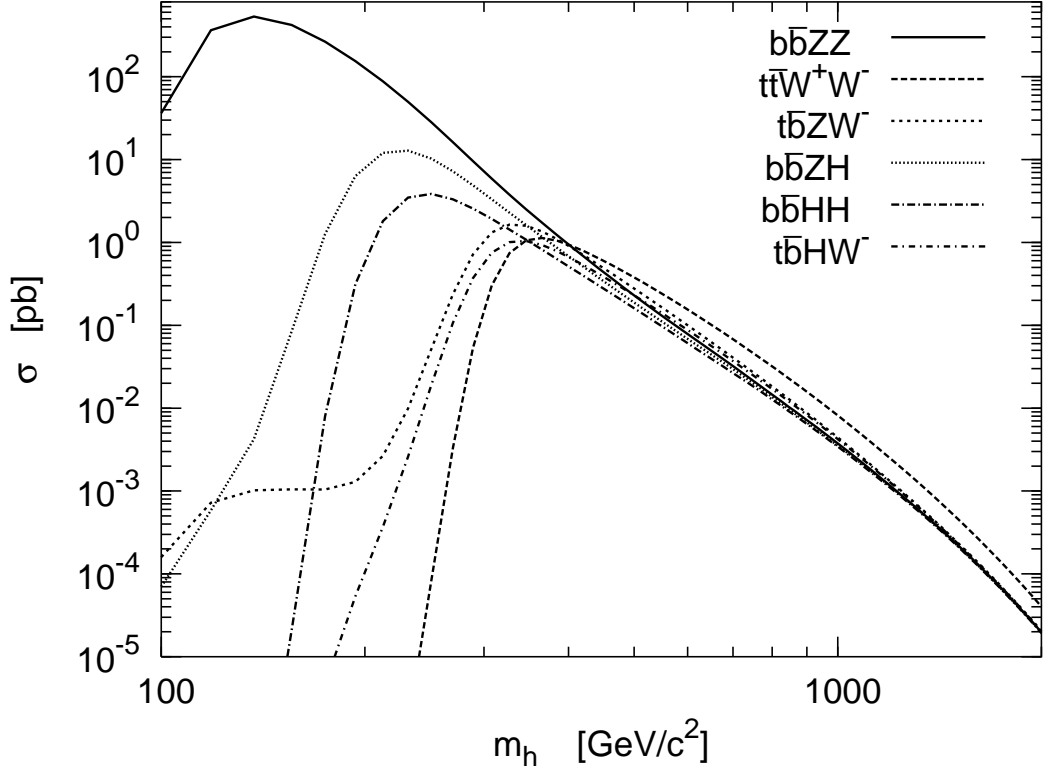}
\caption{\label{fig:xseclhc}Cross sections for $h\bar{h}$ production
and primary decay at the CERN LHC.  The Higgs mass is ${\rm M}_{\rm H}
= 150$ GeV/c$^2$, the new mixing parameter $\xi = 1$, and $Q=2m_h$.}
\end{figure}

The cross sections for $h\bar{h}$ production and (primary $h$ quark)
decay at the LHC are shown in Fig.~\ref{fig:xseclhc}.  In this figure,
the new mixing parameter, $\xi$, is set to $1$, the Higgs mass, $M_H$,
is set to $150$ GeV/c$^2$, and the $h$ quark mass, $m_h$, runs from
$100$ -- $3000$ GeV/c$^2$.  The cross sections for the $h\bar{h}$
primary decay modes at the LHC are roughly two orders of magnitude
larger than the corresponding cross sections at the Tevatron.  Below $300$
GeV/c$^2$, the $b\bar{b}\,{\rm Z}^0{\rm Z}^0$ mode is once again the
dominant primary decay mode of $h\bar{h}$ pair production.  Above an
$h$ quark mass of $300$ GeV/c$^2$, each of the primary decay cross
sections are comparable; however, the decay modes with at least one
charged-current decay of the $h$ quark tend to be larger.

At the LHC the cross sections for each of the three charged-current
primary decay modes are significant, unlike those at the Tevatron.
For an $h$ quark mass of $500$ GeV/c$^2$ and a mixing parameter
$\xi=1$, the LHC should produce $\sim 20\,000$ $h\bar{h}$ events
decaying to the $t\bar{t}\,{\rm W}^+{\rm W}^-$ mode in $100$ fb$^{-1}$
of data.  The charged-current mode of $h\bar{h}$ decay is important
because our parametrization of the \ld\spac (CKM) matrix in the ISVLQ
model predicts a relationship between the size of charged-current and
neutral-current interactions.  As discussed in
Section~\ref{subsec:backgrounds}, the charged-current modes have final
states that are more complicated than those encountered in FCNC decay
modes.  Additional complications are attributed to the decay of the
top quark resulting in an additional ${\rm W}^{\pm}$ boson.  With
additional particles in the final state, dijet mass resolution and the
calculation of signature backgrounds are extremely important
components for an analysis.  Rather than address the charged-current
modes in this paper, we defer this analysis to future research when
the difficulties of the background calculation and details about dijet
mass resolution and $b$-tagging can be better resolved.

In our analysis of $h\bar{h}$ production at the Tevatron [see
Section~\ref{subsec:panddtev}], we found that signatures of the
$b\bar{b}\,{\rm Z}^0{\rm Z}^0$ mode provide the highest $h$ quark mass
reach.  We conclude that in $10$ fb$^{-1}$ of data, $h\bar{h}$ pair
production can be probed up to an $h$ quark mass of $320$ GeV/c$^2$.
At this $h$ quark mass and for a Higgs mass of $150$ GeV/c$^2$, the
cross section to the $b\bar{b}\,{\rm Z}^0{\rm Z}^0$ mode
($\sigma_{bbZZ} = 20$ fb) and the branching ratio to the ``optimal''
signatures combine to produce a handful of signal events with
negligible background.  Based on our study of the $b\bar{b}\,{\rm
  Z}^0{\rm Z}^0$ mode at the Tevatron, we infer a mass reach for this
decay mode at the LHC.  We assume that the branching ratios of the
$b\bar{b}\,{\rm Z}^0{\rm Z}^0$ mode to {\it final-state} signatures and
the detection of these signatures at the LHC are similar to the
Tevatron.  Therefore, at the LHC, the upper limit on the $h$ quark
mass is encountered when approximately $200$ events are produced in
the $b\bar{b}\,{\rm Z}^0{\rm Z}^0$ mode.  The expected integrated
luminosity at the LHC is $100$ fb$^{-1}$; thus one can probe the
$b\bar{b}\,{\rm Z}^0{\rm Z}^0$ cross section down to $2$ fb. This
corresponds to an $h$ quark mass reach of $1100$ GeV/c$^2$ [see
Fig.~\ref{fig:xseclhc}].

%%%%%%%%%%%%%%%%%%%%%%%%%%%%%%%%%%%%%%%%%%%%% Conclusion %%%%%%%%%%%%%%%%%%%%%%%%%%%%%%%%%%%%%%%%%%%%%%%%%
\section{Conclusion}
\label{sec:conclusion}
We have investigated an ${\rm E}_6$-inspired extension of the Standard
Model in which an exotic charge $-1/3$ isosinglet vector-like quark
(denoted $h$) interacts predominantly with the third generation of
quarks.  In this model, the CKM matrix is no longer unitary and it
is replaced by a $3\times 4$ matrix containing new angles and phases.
The loss of CKM unitarity is accompanied by the emergence of {\it
  tree-level} flavor-changing neutral currents mediated by both ${\rm
  Z}^0$ and ${\rm H}^0$ bosons.  Flavor-changing neutral-current
interactions between the $h$ quark and the $b$ quark produce
signatures of $h\bar{h}$ production accessible for detection at hadron
colliders.

At the Fermilab Tevatron, we find that $h$ quark discovery through
pair production is accessible up to an $h$ quark mass of ($270$, $320$)
GeV/c$^2$ in ($1$, $10$) fb$^{-1}$ of data.  Previous $b^{\prime}$
analyses from Run I of the Tevatron
were used to infer that an $h$ quark is currently excluded up to $200$
GeV/c$^2$.  The ($270$, $320$) GeV/c$^2$ mass reach is attainable through the
decay of the $b\bar{b}\, {\rm Z}^0{\rm Z}^0$ mode to the
$b\bar{b}(jj)_Z(l^+ l^-)_Z$ or the $b\bar{b}(jj)_Z\sla{\rm E}_{\rm T}$
signatures.  Furthermore, primary decay of an $h\bar{h}$ pair to the
$b\bar{b}\, {\rm H}^0{\rm Z}^0$ mode provides the opportunity for the
discovery of the Higgs boson in conjunction with $h$ quark discovery.
The viability of the $b\bar{b}\, {\rm H}^0{\rm Z}^0$ mode at the
Tevatron hinges on the branching ratio of the Higgs boson to a $b$
quark pair.  

At the CERN LHC, $h\bar{h}$ pair production is accessible through both
charged-current and neutral-current decays of the $h$ quark.  In $100$
fb$^{-1}$ of data, we find that the $h$ quark mass reach through the
$b\bar{b}\, {\rm Z}^0{\rm Z}^0$ primary decay mode is $1100$
GeV/c$^2$.  To understand the reach of the charged-current primary
decay modes, an analysis of potential signatures, the effect of
detector limitations, and signature backgrounds is necessary.  A
thorough analysis of $h\bar{h}$ charged-current decay modes
($t\bar{b}\, {\rm W}^-{\rm Z}^0$, $t\bar{b}\, {\rm W}^-{\rm H}^0$, and
$t\bar{t}\, {\rm W}^+{\rm W}^-$) at the LHC is deferred to future work.

%\appendix
%\section{}

\begin{acknowledgments}
We thank H. Frisch, C.E.M. Wagner, and D. Morrissey for useful
discussions.  We also thank the University of Chicago CDF group for
computing time.  We are grateful to Stephen Wolbers, Igor
Mandrichenko, and Margaret Greaney from the Computing Division (ISD
department) at the Fermi National Accelerator Laboratory.
\end{acknowledgments}

%\bibliography{e6_paper_revtex}

\begin{thebibliography}{99}
\bibitem{footnote:disfavored}
Though an interesting scenario, a large coupling between the $h$ quark
and $u$ or $c$ quarks appears to be disfavored by bounds discussed in
Section~\ref{subsec:CKMconstraints}.
\bibitem{footnote:reparam}
Technically, for $\xi$ parameters not of ${\mathcal O}(1)$ ({\it e.g.}
$\xi=0.02$), one should reparametrize the $L^d_{34}$ element of the
\ld\spac matrix so that the matrix parametrization is consistent.
\bibitem{footnote:notau}
In the detector, the decay of $\tau$ leptons results in a jet
signature.  In this paper, we restrict our signature analysis to
non-tau leptonic decays of the ${\rm W}^{\pm}$ bosons.
\bibitem{footnote:smearjj}
We use a 10\% dijet mass resolution for both light quark hadronization
and $b$ quark hadronization.  Though optimistic, it is expected that
improved energy flow techniques will attain a resolution commensurate
to this value~\cite{flow}.
\end{thebibliography}

% The Bibliography

\end{document}